\documentstyle[emulateapj,rotate,times]{article}
 
\newcommand{\etal}{et~al.\ }
\newcommand{\PVdblt}{{\rm P}\kern 0.1em{\sc v}~$\lambda\lambda 1117, 1128$}
\newcommand{\CaIIdblt}{{\rm Ca}\kern 0.1em{\sc ii}~$\lambda\lambda 3934, 3969$}
\newcommand{\AlIIIdblt}{{\rm Al}\kern 0.1em{\sc iv}~$\lambda\lambda 1855, 1863$}
\newcommand{\CIVdblt}{{\rm C}\kern 0.1em{\sc iv}~$\lambda\lambda 1548, 1550$}
\newcommand{\MgIIdblt}{{\rm Mg}\kern 0.1em{\sc ii}~$\lambda\lambda 2796, 2803$}
\newcommand{\NVdblt}{{\rm N}\kern 0.1em{\sc v}~$\lambda\lambda 1238, 1242$}  
\newcommand{\SVIdblt}{{\rm S}\kern 0.1em{\sc vi}~$\lambda\lambda 933, 944$} 
\newcommand{\OVIdblt}{{\rm O}\kern 0.1em{\sc vi}~$\lambda\lambda 1031, 1037$} 
\newcommand{\SiIIdblt}{{\rm Si}\kern 0.1em{\sc ii}~$\lambda\lambda 1190, 1193$} 
\newcommand{\SiIVdblt}{{\rm Si}\kern 0.1em{\sc iv}~$\lambda\lambda 1393, 1402$} 
\newcommand{\PV}{\hbox{{\rm P}\kern 0.1em{\sc v}}}
\newcommand{\AlI}{\hbox{{\rm Al}\kern 0.1em{\sc i}}}
\newcommand{\AlII}{\hbox{{\rm Al}\kern 0.1em{\sc ii}}}
\newcommand{\AlIII}{{\hbox{\rm Al}\kern 0.1em{\sc iii}}}
\newcommand{\CaII}{\hbox{{\rm Ca}\kern 0.1em{\sc ii}}}
\newcommand{\CII}{\hbox{{\rm C}\kern 0.1em{\sc ii}}}
\newcommand{\CIIe}{\hbox{{\rm C$^{\ast}$}\kern 0.1em{\sc ii}}}
\newcommand{\CIII}{\hbox{{\rm C}\kern 0.1em{\sc iii}}}
\newcommand{\CIV}{\hbox{{\rm C}\kern 0.1em{\sc iv}}}
\newcommand{\CV}{\hbox{{\rm C}\kern 0.1em{\sc v}}}
\newcommand{\HI}{\hbox{{\rm H}\kern 0.1em{\sc i}}}
\newcommand{\HII}{\hbox{{\rm H}\kern 0.1em{\sc ii}}}
\newcommand{\Lya}{\hbox{{\rm Ly}\kern 0.1em$\alpha$}}
\newcommand{\Lyb}{\hbox{{\rm Ly}\kern 0.1em$\beta$}}
\newcommand{\Lyg}{\hbox{{\rm Ly}\kern 0.1em$\gamma$}}
\newcommand{\Lyd}{\hbox{{\rm Ly}\kern 0.1em$\delta$}}
\newcommand{\Lye}{\hbox{{\rm Ly}\kern 0.1em$\epsilon$}}
\newcommand{\Lyphi}{\hbox{{\rm Ly}\kern 0.1em$\phi$}}
\newcommand{\Lyfive}{\hbox{{\rm Ly}\kern 0.1em$5$}}
\newcommand{\Lysix}{\hbox{{\rm Ly}\kern 0.1em$6$}}
\newcommand{\Lyseven}{\hbox{{\rm Ly}\kern 0.1em$7$}}
\newcommand{\Lyeight}{\hbox{{\rm Ly}\kern 0.1em$8$}}
\newcommand{\Lynine}{\hbox{{\rm Ly}\kern 0.1em$9$}}
\newcommand{\Lyten}{\hbox{{\rm Ly}\kern 0.1em$10$}}
\newcommand{\Lyeleven}{\hbox{{\rm Ly}\kern 0.1em$11$}}
\newcommand{\HeI}{\hbox{{\rm He}\kern 0.1em{\sc i}}}
\newcommand{\HeII}{\hbox{{\rm He}\kern 0.1em{\sc ii}}}
\newcommand{\FeI}{\hbox{{\rm Fe}\kern 0.1em{\sc i}}}
\newcommand{\FeII}{\hbox{{\rm Fe}\kern 0.1em{\sc ii}}}
\newcommand{\FeIII}{\hbox{{\rm Fe}\kern 0.1em{\sc iii}}}
\newcommand{\MnII}{\hbox{{\rm Mn}\kern 0.1em{\sc ii}}}
\newcommand{\MgI}{\hbox{{\rm Mg}\kern 0.1em{\sc i}}}
\newcommand{\MgII}{\hbox{{\rm Mg}\kern 0.1em{\sc ii}}}
\newcommand{\MgIII}{\hbox{{\rm Mg}\kern 0.1em{\sc iii}}}
\newcommand{\NI}{\hbox{{\rm N}\kern 0.1em{\sc i}}}
\newcommand{\NII}{\hbox{{\rm N}\kern 0.1em{\sc ii}}}
\newcommand{\NIII}{\hbox{{\rm N}\kern 0.1em{\sc iii}}}
\newcommand{\NV}{\hbox{{\rm N}\kern 0.1em{\sc v}}}
\newcommand{\OVI}{\hbox{{\rm O}\kern 0.1em{\sc vi}}}
\newcommand{\OI}{\hbox{{\rm O}\kern 0.1em{\sc i}}}
\newcommand{\OII}{\hbox{[{\rm O}\kern 0.1em{\sc ii}]}}
\newcommand{\OIV}{\hbox{{\rm O}\kern 0.1em{\sc iv}]}}
\newcommand{\SI}{{\rm S}\kern 0.1em{\sc i}}
\newcommand{\SIV}{{\rm S}\kern 0.1em{\sc iv}}
\newcommand{\SVI}{{\rm S}\kern 0.1em{\sc vi}}
\newcommand{\SiI}{\hbox{{\rm Si}\kern 0.1em{\sc i}}}
\newcommand{\SiII}{\hbox{{\rm Si}\kern 0.1em{\sc ii}}}
\newcommand{\SiIII}{\hbox{{\rm Si}\kern 0.1em{\sc iii}}}
\newcommand{\SiIV}{\hbox{{\rm Si}\kern 0.1em{\sc iv}}}
\newcommand{\SII}{\hbox{{\rm S}\kern 0.1em{\sc ii}}}
\newcommand{\SIII}{\hbox{{\rm S}\kern 0.1em{\sc iii}}}
\newcommand{\NaI}{\hbox{{\rm Na}\kern 0.1em{\sc i}}}
\newcommand{\TiII}{\hbox{{\rm Ti}\kern 0.1em{\sc ii}}}
\newcommand{\kms}{\hbox{km~s$^{-1}$}}
\newcommand{\cmsq}{\hbox{cm$^{-2}$}}

\tighten
\begin{document}
 
\slugcomment{The Astrophysical Journal, {\rm submitted}}
 
\lefthead{CHURCHILL ET~AL.}
\righthead{{\MgII} ABSORPTION SELECTED GALAXIES}


\title{Low and High Ionization Absorption Properties of {\MgII}
Absorption--Selected Galaxies at Intermediate Redshifts. II. Taxonomy,
Kinematics, and Galaxies\altaffilmark{1,2}}

\author{Christopher~W.~Churchill\altaffilmark{3}, Richard~R.~Mellon,
Jane~C.~Charlton\altaffilmark{4}}

\affil{The Pennsylvania State University, University Park, PA 16802}

\author{Buell~T.~Jannuzi}
\affil{National Optical Astronomy Observatories, Tucson, AZ 85719}

\author{Sofia~Kirhakos}
\affil{Institute for Advanced Study, Princeton, NJ 08544}

\author{Charles~C.~Steidel\altaffilmark{5}}
\affil{California Institute of Technology, Palomar Observatories,
Pasadena, CA 91125}

\and

\author{Donald~P.~Schneider}
\affil{The Pennsylvania State University, University Park, PA 16802}

\altaffiltext{1}{Based in part on observations obtained at the
W.~M. Keck Observatory, which is operated as a scientific partnership
among Caltech, the University of California, and NASA. The Observatory
was made possible by the generous financial support of the W. M. Keck
Foundation.}
\altaffiltext{2}{Based in part on observations obtained with the
NASA/ESA {\it Hubble Space Telescope}, which is operated by the STScI
for the Association of Universities for Research in Astronomy, Inc.,
under NASA contract NAS5--26555.}
\altaffiltext{3}{Visiting Astronomer at the W.~M. Keck Observatory}
\altaffiltext{4}{Center for Gravitational Physics and Geometry}
\altaffiltext{5}{NSF Young Investigator}

\begin{abstract}
We examine a sample of 45 {\MgII} absorption--selected systems
over the redshift range 0.4 to 1.4 in order to better understand the
range of physical conditions present in the interstellar and halo gas
associated with intermediate redshift galaxies.
{\MgII} and {\FeII} absorption profiles were observed at a resolution
of $\simeq 6$~{\kms} with HIRES/Keck.
{\Lya} and {\CIV} data were measured in FOS spectra obtained from the
{\it Hubble Space Telescope\/} archive (resolution $\simeq 230$~{\kms}).
We perform a multivariate analysis of $W_{r}({\MgII})$,
$W_{r}({\FeII})$, $W_{r}({\CIV})$ and $W_{r}({\Lya})$ (rest--frame
equivalent widths) and the {\MgII} kinematic spread. 
There is a large range of high--to--low ionization properties and 
kinematics in intermediate redshift absorbers, that we find can
be organized into five categories: ``Classic'', ``{\CIV}--deficient'',
``Single/Weak'', ``Double'', and ``DLA/{\HI}--Rich''.
These categories arise, in part, because there is a strong connection
between low--ionization kinematics and the location of an absorber on
the $W_{r}({\CIV})$--$W_{r}({\MgII})$ plane.
Using photoionization modeling, we infer that in most absorbers a
significant fraction of the {\CIV} arises in a phase separate from
that giving rise to the {\MgII}.
We show that many of the {\CIV} profiles are resolved in the FOS
spectra due to velocity structure in the {\CIV} gas.
For 16 systems, the galaxy $M_K$, $M_B$, $B-K$, and impact parameters
are measured.
We compare the available absorption--line properties (taken from
Churchill \etal 1999, Paper I) to the galaxy properties but find no
significant (greater than $3~\sigma$) correlations, although several
suggestive trends are apparent.
We compare the locations of our intermediate redshift absorbers on the
$W_{r}({\CIV})$--$W_{r}({\MgII})$ plane with those of lower and higher
redshift data taken from the literature and find evidence for
evolution that is connected with the {\MgII} kinematics seen in
HIRES/Keck profiles of {\MgII} at $z > 1.4$.
We discuss the potential of using the above categorizations of
absorbers to understand the evolution in the underlying physical
processes giving rise to the gas and governing its ionization phases
and kinematics.
We also discuss how the observed absorbing gas evolution has
interesting parallels with scenarios of galaxy evolution in which
mergers and the accretion of ``proto--galactic clumps'' govern the gas
physics and provide reservoirs for elevated star formation rates at
high redshift.
At intermediate and lower redshifts, the galaxy gaseous components and
star formation rates may become interdependent and self--regulatory
such that, at $z \leq 1$, the kinematics and balance of high and low
ionization gas may be related to the presence of star forming regions
in the host galaxy.

\end{abstract}


\section{Introduction}
\label{sec:intro}
 
It is well known that individual galaxies fall into a fairly
clear--cut morphological classification scheme, as first proposed by
Hubble (1936\nocite{hubble}).
Over the decades, Hubble's scheme has been expanded upon and exploited
for quantifying galactic evolution, galactic stellar populations and
star formation efficiencies, the relative gas--phase component to
stellar component in galaxies, and the effects of environment on
galaxy formation.  
Similarly, one of the central motivations for studying intervening
quasar absorption lines, especially those selected by the presence of
metal lines, is that they also provide insights into galactic
evolution, not of the stars and stellar dynamics, but of the chemical,
ionization, and kinematic conditions of interstellar and halo gas.

Historically, absorption line systems have been classified in a
taxonomic system by their gas cross sections.  
Categorized by increasing {\HI} column densities are the {\Lya} forest
clouds, with $N({\HI)} \leq 10^{15}$~{\cmsq}, the sub--Lyman limit
systems, having $N({\HI)} \sim 10^{16}$~{\cmsq}, the Lyman limit break
systems, with $N({\HI)} \geq 10^{17.3}$~{\cmsq}, and the damped {\Lya}
systems, having $N({\HI)} \geq 10^{20.3}$~{\cmsq}.
Metal--line systems are either selected by the presence of strong
{\MgIIdblt} absorption or strong {\CIVdblt} absorption.
 
What has not been explored, however, is the taxonomy of absorption
line systems when the {\HI}, {\MgII}, and {\CIV} absorption strengths
and the gas kinematics are equally considered. 
Ultimately, categorizing the relationships between several different
absorption properties may provide clues central to understanding the
different physical natures of the various types of systems.
 
Furthermore, while our empirical knowledge of the absorption
strengths, kinematics, and physical extent of galaxies selected by
{\MgII} absorption has steadily progressed over the last decade
(e.g.\ \cite{bb91}; \cite{lb90}, 1992\nocite{lb92}; \cite{bergeron92};
\cite{lebrun93}; \cite{sdp94}; \cite{csv96}; \cite{guillemin}), little
is {\it directly\/} known about {\HI} absorption and higher ionization
absorption (esp.\ {\SiIV}, {\CIV}, {\NV}, and {\OVI}) in these
galaxies.

Arguably, the {\MgII}--selected systems are ideally
suited for a taxonomic study of absorption systems because:
 
(1) those with $W_{r}({\MgII}) \geq 0.3$~{\AA} are known to be
directly associated with galaxies (\cite{bb91}; Steidel, Dickinson, \&
Persson 1994\nocite{sdp94}; \cite{csv96}) and/or sub--galactic
metal--enriched environments (\cite{yanny92}; \cite{yannyyork92}).
{\it HST\/} imaging has revealed that these galaxies have a wide
variety of line--of--sight orientations and ``normal'' morphologies
(see \cite{3c336}; \cite{steidel98}).
Since magnesium is an $\alpha$--group element yielded by Type II
supernovae, it is expected that the association with galaxies will
hold to the highest redshifts.

(2) they arise in structures having a five decade range of {\HI} column
densities, including sub--Lyman limit systems (\cite{weak};
Churchill \etal 2000, hereafter Paper~I\nocite{paperI}), Lyman limit
systems (e.g.\ \cite{ss92}; Paper~I\nocite{paperI}), and damped
{\Lya} systems  (e.g.\ \cite{lebrun97}; \cite{rao98}; \cite{boisse})
which means that a large range of galactic environments will be
sampled.

(3) for $z < 2.2$ their statistical properties and for $z < 1.4$ their
kinematic properties have been thoroughly documented 
(e.g.\ \cite{ltw87}; \cite{sbs88}; \cite{ss92}; \cite{pb90};
\cite{ss92}; \cite{thesis}; \cite{weak}) which means that the low
redshift database is already in place and can be used for evolution
studies when higher redshift data are obtained.

(4) they are seen to give rise to a range of {\CIV} and other high
ionization absorption in UV space--based spectra  (\cite{bergeronKP};
Paper~I\nocite{paperI}), which means that the more
general ionization conditions can be studied in detail (e.g.\
\cite{q1206}).

In this paper, we study the combined available data on a sample of 45
{\MgII} absorption--selected systems at redshifts $\sim 0.4$--$1.4$.
We focus on the {\MgII} and {\FeII} absorption strengths and
kinematics, the {\Lya} and {\CIV} absorption strengths
(Paper~I\nocite{paperI}), and the ground--based derived galaxy
luminosities, colors, and impact parameters.
We perform a multivariate analysis (\cite{babu}; \cite{johnson}) on
the absorption line data in order to objectively quantify systematic
differences between and/or groupings of the high ionization and
neutral hydrogen absorption properties of  {\MgII} selected systems.

The paper is organized as follows: 
In \S~\ref{sec:data}, we describe the data, the sample selection, and
the data analysis.
In \S~\ref{sec:classes}, we investigate the variation in the high and
low ionization properties of intermediate redshift {\MgII} absorbers
using multivariate analysis.
We introduce a taxonomy that serves as an objective guide for
classifying the variations in {\MgII} absorber properties.
In \S~\ref{sec:multiphase}, we investigate the overall ionization
and kinematic conditions.
In \S~\ref{sec:galaxies} we investigate the relationship between the
absorption properties and host galaxy properties and in
\S~\ref{sec:discussion} offer a speculative discussion on the
relationships between the absorber classes and their possible
evolution.
In \S~\ref{sec:summary} we summarize the main points of this work.

\section{The Data: Sample Selection and Analysis}
\label{sec:data}

We targeted quasar absorption line systems at intermediate redshifts
that have been discovered by the presence of a {\MgIIdblt} doublet
(e.g.\ \cite{ltw87}; \cite{sbs88}; \cite{ss92}).
We also include in our study the ``weak'' systems [those with
$W_{r}({\MgII}) < 0.3$~{\AA}], which are more numerous in
their redshift path density  (\cite{weak}).
A detailed account of the sample selection is given in
Paper~I\nocite{paperI}, but we briefly outline the sample properties
here.

The {\MgII} absorbers were selected from a high resolution ($\sim
6$~{\kms}) survey (\cite{thesis}) using the HIRES
spectrograph (\cite{vogt94}) on the Keck~I telescope.
For 45 of these systems, additional wavelength coverage in the
ultraviolet was available in the {\it HST\/}/FOS archive and the
database compiled by the {\it HST\/} QSO Absorption Line Key Project
(\cite{cat1}; \cite{cat2}; \cite{cat3}).

The resulting database of {\MgII} absorbing systems and their
rest--frame equivalent widths are listed in Table~\ref{tab:lineprops}.
The first three columns are the quasar name, the absorber redshift,
and the {\MgII} $\lambda 2796$ rest--frame equivalent width,
$W_{r}({\MgII})$.
The fourth and fifth columns list the {\MgII} kinematic spread,
$\omega _{v}$ (see Equation~\ref{eq:omegav}), and the kinematic
composition, i.e.\ the number of Voigt Profile components (see
\S~\ref{sec:kinematics}). 
Columns six, seven, and eight list the {\FeII}, {\CIV}, and {\Lya}
rest--frame equivalent widths.
Column nine gives the various subsamples used for our analyses.
Column ten lists the taxonomic ``class'' of each system (explained in
\S~\ref{sec:clusteringdiscussion}).

Full details of the data analysis are given in Paper~I\nocite{paperI},
including the continuum fitting, the line finding, the equivalent
width measurements, the establishment of a redshift zero point for the
FOS spectra, the procedure for line identifications, and the
techniques employed to measure Lyman limit breaks, when present.

As shown in Figure 2 of Paper~I\nocite{paperI}, the {\MgII} equivalent
width distribution is consistent with that of an unbiased sample of
absorbers for $W_{r}({\MgII}) \leq 1.3$~{\AA} (\cite{weak}).
There are five systems with $W_{r}({\MgII}) > 1.3$~{\AA}, which may
appear as a bias toward strong systems.  
When this may be a concern for our analysis, we discuss this
possibility.

With regard to detection sensitivity, the sample is 72\% complete to a
$W_{r}({\MgII})$ rest--frame detection threshold of $0.02$~{\AA} and
93\% complete to a $0.03$~{\AA} threshold (Figure 3 of
Paper~I\nocite{paperI}).
In column nine of Table~\ref{tab:lineprops}, we have designated those
systems having rest--frame detection thresholds greater than 
$0.03$~{\AA} as Sample A.
For these absorbers, unresolved {\MgII} absorption features with 
$W_{r}({\MgII}) \leq 0.03$~{\AA} cannot be detected.
The other subsample designations, ``CA'' and IC'', are explained in
\S\S~\ref{sec:casample} and \ref{sec:multiphase}, respectively.

\subsection{Galaxy Sample}

For roughly 60 {\MgII} absorbers having $W_{r}({\MgII}) \geq
0.3$~{\AA}, Steidel \etal (1994\nocite{sdp94}) identified associated galaxies.
Only 16 of the 45 {\MgII} systems presented here, of which 21 have
$W_{r}({\MgII}) \geq 0.3$~{\AA}, have confirmed galaxy counterparts
(not all fields studied here have been imaged).

The galaxy properties were obtained from broad--band $g (4900/700)$,
$\Re (6930/1500)$, $i(8000/1450)$ and $K$--band images of the QSO
fields and their redshifts were spectroscopically verified to be
coincident (within $\sim 100$~{\kms}) with those of the {\MgII}
absorbers. 
PSF subtraction of the quasar was performed for all fields, enabling
small impact parameter galaxies to be identified.
Further details of the imaging and spectroscopic observations are
described in Steidel \etal (1994\nocite{sdp94}).

The galaxy properties are presented in Table~\ref{tab:galprops};
the columns from left to right are the quasar field, the galaxy
redshift, the absolute $B$ magnitude, $M_B$, the absolute $K$
magnitude, $M_K$, the de--reddened $B-K$ color, the galaxy--quasar
impact parameter in $h^{-1}$kpc ($H_0 = 100$~{\kms}~Mpc$^{-1}$, $q_0
= 0.05$), and a reference if previously published.
In a few cases, morphology information is available from published
{\it HST\/} images (\cite{lebrun97}; \cite{3c336}; \cite{steidel98}).

It is always possible that a galaxy is misidentified, that another
galaxy or more than one galaxy is giving rise to the absorption.
Possible selection effects due to misidentifications and
incompleteness of galaxy redshifts in the individual quasar fields are
discussed by Steidel \etal (1994\nocite{sdp94}) and examined in detail
by Charlton \& Churchill (1996\nocite{cc96}). 

\subsection{Data Analysis: Kinematics}
\label{sec:kinematics}

To measure the kinematic ``spread'' of the low ionization gas directly
from the flux values, we use the second velocity moment of optical
depth across the {\MgII} $\lambda 2796$ profile, defined as,
\begin{equation}
\omega _{v} = 
\left\{
\frac { \sum \tau_{a}(v_{i}) v_{i}^{2} \Delta v_{i} }
{ \sum \tau_{a}(v_{i}) \Delta v_{i} } 
\right\} ^{1/2},
\label{eq:omegav}
\end{equation}
where $\tau_{a}(v_{i}) = \ln [I_{c}(v_{i})/I(v_{i})]$ is the apparent
optical depth measured directly from the flux values
(\cite{savage_aod}) at velocity $v_{i}$, 
$\Delta v_{i} = (v_{i-1}-v_{i+1})/2$, $v_{i} = c (\lambda _{i}/\lambda
_{obs} - 1)$, and $\lambda _{obs} = 2796.352(1+z_{\rm abs})$.
The sums are performed only over velocity intervals in which
absorption features have been detected at the $5~\sigma$ level,
thus eliminating terms consistent with noise (see
Paper~I\nocite{paperI} for details on feature detection).
The quantity measured by $\omega _{v}$ is similar to the standard
deviation measured from a normal distribution.
The uncertainty in $\omega _{v}$ is obtained from simple error
propagation (\cite{thesis}; also see \cite{kenspaper}).
The velocity zero point of each system is set at the median wavelength
of the apparent optical depth distribution of the {\MgII} $\lambda
2796$ profile.

The value of $\omega _{v}$ depends upon the velocity squared, and thus
its value is sensitive to the presence of very weak {\MgII}
absorption at large velocity.
Therefore, $\omega _{v}$ is sensitive to the equivalent width detection
threshold of the HIRES spectra.
We are 93\% complete to a $5~\sigma$ rest--frame equivalent width
detection threshold of $0.03$~{\AA} for the full sample of 45 systems
(see Figure~3 from Paper~I\nocite{paperI}).
To enforce uniform evaluation of the kinematic spread for the full
sample, we omitted features with $W_{r}({\MgII}) < 0.03$~{\AA} from
the computation of $\omega_{v}$.
Six systems required this censorship step; they are footnoted in
Table~\ref{tab:lineprops}.
The maximum change in  $\omega _{v}$ from its uncensored value was a
$-20$\% difference, except for one of the absorbers (PKS~$0454+039$ at
$z=1.1532$), which was changed by a $-40$\% difference.

To measure the kinematic ``composition'' of the low ionization gas, we
use Voigt profile fitting.
We use our own program, MINFIT (\cite{thesis}), which performs a $\chi
^{2}$ minimization using the NETLIB--{\sc slatec} routine {\sc dnls1}
(\cite{more78}).
In fully saturated profiles, the number of clouds is often
underestimated by a factor of a few (\cite{thesis}). 
A corollary is that the cloud column densities, $b$ parameters, and
velocities are correlative with the number of clouds. 
We return to this issue in \S~\ref{sec:dataresults}.


\section{Characterizing {\MgII} Absorber Properties}
\label{sec:classes}
\label{sec:casample}

We have constructed a subsample of 30 systems, Sample CA (CA = Cluster
Analysis), in order to examine variations and possible trends between
the absorption properties listed in Table~\ref{tab:lineprops}.
All Sample CA members have {\it measured\/} $W_{r}({\Lya})$,
$W_{r}({\MgII})$, and $\omega _{v}$ and have {\it measurements of or
limits on\/} {\FeII} and {\CIV}.
Systems with upper limits on $W_{r}({\FeII})$ were included in the
sample because these limits are already capable of demonstrating a
paucity of {\FeII} absorption (only seven Sample CA systems have
{\FeII} limits).
Small $W_{r}({\MgII})$ systems with upper limits on {\CIV} were
excluded from the sample because the limits were not always stringent.
However, multiple cloud systems with limits on $W_{r}({\CIV})$ and
larger $W_{r}({\MgII})$ are included.
The members of Sample CA are given in column nine of
Table~\ref{tab:lineprops}.

\begin{figure*}[t]
\plotfiddle{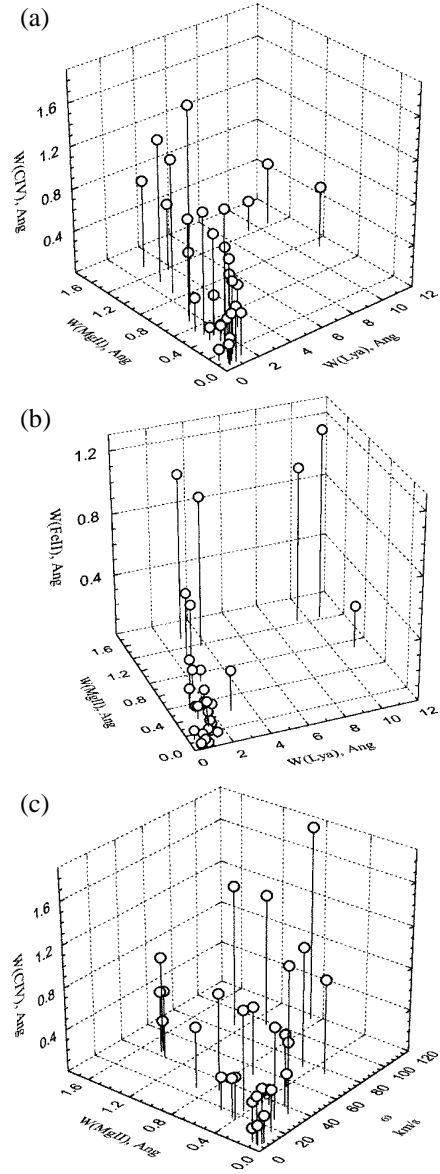}{6.in}{0}{70}{70}{0}{0}
\caption[fig1.ps]
{Three dimensional scatter plots of 
(a) $W_{r}({\MgII})$ vs.\ $W_{r}({\Lya})$ vs.\ $W_{r}({\CIV})$,
(b) $W_{r}({\MgII})$ vs.\ $W_{r}({\Lya})$ vs.\ $W_{r}({\FeII})$,
(c) $W_{r}({\MgII})$ vs.\ $\omega _{v}$ vs.\ $W_{r}({\CIV})$.
Upper limits on {\FeII} and {\CIV}, of which a few are present near
the origin, are plotted but not delineated.  Vertical lines intersect
the $x$--$y$ values (right handed coordinate system) in the $x$--$y$
plane.  The $z$ value of a datum can be obtained by projecting a
line from the point to the background mesh in a direction parallel to
the constant $z$ lines in the mesh. \label{fig:threed}}
\end{figure*}

In Figure~\ref{fig:threed}, we present three dimensional plots
of 
(a) $W_{r}({\MgII})$ vs.\ $W_{r}({\Lya})$ vs.\ $W_{r}({\CIV})$,
(b) $W_{r}({\MgII})$ vs.\ $W_{r}({\Lya})$ vs.\ $W_{r}({\FeII})$, and
(c) $W_{r}({\MgII})$ vs.\ $\omega _{v}$ vs.\ $W_{r}({\CIV})$ for
Sample CA.
Note that distribution of equivalent widths are not random; there is a
clear trend for $W_{r}({\MgII})$ to increase with $W_{r}({\Lya})$, for
example.
On the other hand, it is clear that there are significant spreads, or
variations in the absorption strengths.
In the case of $W_{r}({\FeII})$, we see a very large range of values
(Figure~\ref{fig:threed}$b$) that trace  $W_{r}({\Lya})$, and to a
lesser extent  $W_{r}({\MgII})$.
$W_{r}({\CIV})$ exhibits a significant spread for a given
$W_{r}({\MgII})$--$W_{r}({\Lya})$ locus (Figure~\ref{fig:threed}$a$).
Moreover, note the groupings of $W_{r}({\MgII})$, $\omega
_{v}$, and $W_{r}({\CIV})$ (Figure~\ref{fig:threed}$c$).

From an empirical point of view, the qualitative appearance
of groupings in Figure~\ref{fig:threed} would suggest that
{\MgII}--selected systems can be further categorized (quantitatively)
by their absorption properties.
However, our sample is small, having only 30 data points, and it
is not clear whether the apparent groupings would statistically be
present in other ``realizations'' of the data\footnote{Monte Carlo
realizations of our sample could shed light on this issue if, and only
if, we had {\it a priori\/} knowledge of the distribution functions
for all the properties being studied.}.
This concern is best addressed by using a multivariate analysis in
which all available absorption properties are incorporated
simultaneously (multidimensional version of Figure~\ref{fig:threed}
with one dimension for each property).

When the {\Lya}, {\FeII}, and {\CIV} and {\MgII} kinematics are 
simultanously considered, do we find various ``classes'' of {\MgII}
absorbers?
In other words, can we quantitatively describe both the variations
and the trends by considering all the absorption line data presented
in Table~\ref{tab:lineprops}?
To address these questions, we applied ``Tree Clustering'' and
``$K$--means Clustering'' analysis to sample CA.

\subsection{A Multivariate Analysis}

Multivariate clustering analysis algorithms are designed to organize 
data of many variables into catagories so that natural groupings of
the data can be examined in a completely unbiased manner (i.e.\ no
model is imposed upon the data).
Full details of these techniques can be found elsewhere 
(e.g.\ Johnson \& Wichern 1992\nocite{johnson}; Babu \& Feigelson
1996\nocite{babu}); below, we provide limited background material.
We used the STATISTICA software package ({\it www.statsoft.com\/}).

\subsubsection{Tree Cluster Analysis}
\label{sec:treecluster}

In tree cluster analysis, the data occupy a multi--dimensional
space, one dimension for each measured absorption property.
Clustering algorithms compute the ``distances'' between each pair 
of points (absorbers) and then amalgamate them into clusters.
We note that tree cluster analysis is not subject to significance
testing because the result itself is the most significant solution
under the assumption of no {\it a priori\/} hypothesis regarding the
data.

The amalgamation process begins with each absorber in a unique class by
itself and then proceeds by relaxing the criterion of uniqueness in
subsequent steps.
With each step, the algorithm amalgamates larger and larger clusters
of increasingly disimilar properties, until in the final step all
absorbers are joined together in a single class.
Graphically, clusters appear as distinct ``branches'' in a
hierarchical tree, with similarities linked at nodes.
We used Euclidean distances [i.e.\ 
$(\sum_{i} [x_{i}-y_{i}]^{2})^{1/2}$] and Ward's method (Ward
1963\nocite{ward63}) for the amalgamation algorithm.
Ward's method minimizes the sum of squares of the distances between
clusters.
The combination of Euclidean distances and Ward's amalgamation rule
uses an analysis of variances approach for which $N(0,1)$
standardization of the data is optimal.
With this standardization, each variable is baselined and scaled to
have a zero mean and unity standard deviation.
$N(0,1)$ standardization ensures that the distance between any pair of
points in multidimensional space is not biased by a large dynamic
range in one or more of the variables (dimensions).

\subsubsection{$K$--means Clustering}
\label{sec:k-means}

In $K$--means clustering analysis, we begin with the assumption of a
set number of clusters and the algorithm finds the most significant,
or distinct, clusters possible.
Starting with $K$ {\it random\/} clusters, the algorithm moves points
(absorbers) between clusters until both the variability within
clusters is minimized and the variability between clusters is
maximized.
We find that $K=5$ is the highest number of clusters allowed such that
all clusters are considered significant, based upon MANOVA tests
(\cite{johnson}; also see \cite{mukherjee}).
For five clusters, the MANOVA  probabilities, $p$, for accepting a
result as valid, (i.e.\ representative of a unique population) are
highly significant. 
We obtained $p < 10^{-5}$ for each cluster.


\subsection{Results}
\label{sec:clusteringdiscussion}

In Figure~\ref{fig:treecluster}, we present a dendrogram, the tree
diagram showing the results of our cluster analysis.
Along the bottom horizontal axis are the individual {\MgII} systems
identified by quasar and absorber redshift.
The vertical axis is the linkage distance, $LD$, the distance between
clusters.
The larger the value of $LD$ between two branches, the less related
are the objects on each branch.

At $LD \sim 16$ there is a natural grouping into three clusters.
These three clusters are {\it predominantly\/} distinguished by very
strong {\Lya} [right branch, labeled ``DLA/{\HI}--rich''], weaker
{\CIV} [center branch, labeled ``{\CIV}--weak''], and intermediate to
stronger {\CIV} [left branch, labeled ``{\CIV}--strong''] absorption. 

Further insight into the physical differences in the absorbers can be
obtained if we adopt a ``less significant'' linkage distance.
At $LD\sim 4$ there is a secondary grouping into six clusters.
The {\CIV}--strong and {\CIV}--weak clusters have each been separated
by variations in {\MgII} strengths and kinematics and the
DLA/{\HI}--Rich cluster has been separated by a spread in the {\Lya}
strengths.
However, we treat the DLA/{\HI}--Rich systems as a single cluster 
because the MANOVA significance test from a $K$--means clustering
analysis indicates that the further splitting of the DLA/{\HI}--Rich
cluster is not justified.
On Figure~\ref{fig:treecluster}, we have labeled each of these five
clusters as ``Classic'', ``Double'', ``Single/Weak'', ``{\CIV}
deficient'', and ``DLA/{\HI}--Rich''. 

It is in a diagram of the $K$--cluster means,
Figure~\ref{fig:k-means}, that the properties distinguishing each
cluster become apparent.
Across the horizontal axis of Figure~\ref{fig:k-means} are the
absorption properties; from left to right they are the {\MgII},
{\Lya}, {\CIV}, and {\FeII} absorption ``strengths'', and the
``strength'' of the kinematic spread. 
Recall that the data have been $N(0,1)$ standardized.
Thus, for each given absorption property, the mean value for {\it all \/}
systems is zero and the vertical axis is in units of standard
deviations.
Each cluster is represented by a different data point type:
Classic (solid circle); Double (solid square); Single/Weak (open
circle); {\CIV} deficient (solid triangle); and DLA/{\HI}--Rich (solid
pentagon).
These classes are also connected by unique line types (i.e.\
dash--dot, solid, etc.); it is important to not only compare a given
property across clusters, but to compare how {\it all\/} properties
are segregated by cluster.
We now briefly discuss some distinguishing features of each absorber
class.

\subsubsection{Classic Systems}

This class appears to have what might be thought of as
``typical'', or non--extreme properties. 
As seen in Figure~\ref{fig:k-means}, this class is characterized by
having {\MgII}, {\Lya}, {\FeII}, and {\CIV} equivalent widths, and a
kinematic spread within $0.5~\sigma$ of the respective means for the
overall sample.

We call these systems ``Classic'' because they can be thought to
represent the most common type of {\MgII} absorber observable in
earlier generation, low resolution surveys (e.g.\ \cite{ss92};
\cite{ltw87}).
To the sensitivities of these surveys, it was found that virtually
all {\MgII} absorption--selected systems also had {\CIV} absorption.

\subsubsection{{\CIV} ``Deficient'' Systems}

The {\CIV}--deficient systems 
have {\MgII}, {\Lya}, and {\FeII} properties identical to the Classics
and constitute a similar fraction of the overall {\MgII} absorber
population.
Their distinguishing properties are significantly lower {\CIV}
absorption strengths.

\subsubsection{Double Systems}

The Doubles are set apart from the Classics by having
at least twice the {\MgII} and {\CIV} absorption strengths and
{\MgII} kinematics.
Since the mean {\MgII} strength for Doubles is in the same regime
as of the DLA/{\HI}--Rich systems,
it is the {\CIV} strengths and extreme {\MgII} kinematics that set
this class apart from the others.
Doubles also have {\Lya} and {\FeII} absorption strengths that are
systematically greater than those of the Classics.

\subsubsection{Single/Weak Systems}

The Single/Weak class is defined foremost by very small
{\MgII} strengths and kinematics.
In fact, with the exception of two systems, all are single, narrow
clouds (often unresolved in the HIRES spectra).

The Single/Weak systems are underrepresented in Sample CA, mostly
because seven of the systems in our database have no measured
{\Lya}.
If these seven systems were included with their {\CIV} limits artificially
treated as detections, the mean {\CIV} for this class would drop to
$-1.1$ (this being an upper limit) in the $K$--means cluster diagram
(Figure~\ref{fig:k-means}).
Thus, there is a spread in the distribution of {\CIV} strengths
associated with Single/Weak systems.
The true mean {\FeII} is also probably lower than that shown on 
Figure~\ref{fig:k-means} because the limits on {\FeII} in the HIRES
spectra scatter about the measured mean.
Thus, higher sensitivity spectra would reduce the mean {\FeII} for this
class and could reveal a spread in {\FeII} strengths.
Only for the Single/Weak class, for which many of the properties were
below our detection thresholds, are the above issues pertinent.

\subsubsection{DLA/{\HI}--Rich Systems}

The DLA/{\HI}--Rich class has very strong {\MgII},
and extremely strong {\Lya} and {\FeII}, relative to the overall
{\MgII} absorber population.
They have {\CIV} strengths below that of the Classics and {\MgII}
kinematics typical of the Classics and {\CIV}--deficient absorbers.
Their {\MgI} and {\FeII} strengths are five to ten times greater than
those of the typical Classic absorber. 
Unlike the elevated {\Lya} strengths, which are due to broad damping
wings, the {\MgI} and {\FeII} strengths are driven by the
line--of--sight kinematic spreads of the gas for multiple, saturated
components.
For the intermediate ionization species, {\SiII}, {\AlII}, {\CII} and
{\SiIII}, the strengths are somewhat greater than those of the
Classics, on average.
The higher ionization species, {\SiIV}, {\NV} and {\OVI} (to
the extent the latter two have been measured), have strengths
consistent with those of the Classics absorbers.

\subsection{Robustness of the Clustering Analysis}

Two concerns regarding the clustering analysis are: 
(1) possible biasing due to an slight, but apparent, overabundance of
absorbers with $W_{r}({\MgII}) > 1.3$~{\AA}, and 
(2) the exclusion of the ``weak'' systems for which only upper limits
on the {\CIV} equivalent widths were measured.

The systems with $W_{r}({\MgII}) > 1.3$~{\AA} are all members of the
DLA/{\HI}--Rich class.  
Of these, four are bonified DLAs with $N({\HI}) \geq 2\times
10^{20}$~{\cmsq}, whereas two have column densities below this
classical threshold (i.e.\ they are {\HI}--Rich systems).
In an unbiased survey for $z < 1.65$ DLAs, using {\MgII} absorption
as a selection method, Rao \& Turnshek (1999\nocite{rao99}) found that
14\% of the {\MgII} systems with  $W_{r}({\MgII}) > 0.3$~{\AA} are
DLAs.
In our sample, 22 systems have $W_{r}({\MgII}) > 0.3$~{\AA}; thus,
18\% of our sample is comprised of DLAs.
This is not inconsistent with the unbiased results of Rao \& Turnshek.

The presence of a DLA/{\HI}--Rich class in our sample is not due to
the slight bias (overabundance) of the larger $W({\MgII})$;
it is mostly defined by large {\HI} and {\FeII} equivalent widths.
In fact, one DLA has $W_{r}({\MgII}) = 0.9$~{\AA} ($z=0.5764$
toward Q~$0117+213$), and was classified as DLA/{\HI}--Rich in spite
of the fact that this {\MgII} equivalent width is well below
$1.3$~{\AA}.
In other words, even if the {\MgII} equivalent widths of the
DLA/{\HI}--Rich systems were quite small, the class would be unchanged
in the cluster analysis due to the very strong {\HI} and {\FeII}
strengths.
This is clearly shown in Figure~\ref{fig:k-means}.

To investigate whether membership in a given class (especially the
Single/Weak class) may be sensitive to the inclusion of systems for
which only an upper limit is available for $W_{r}({\CIV})$, we ran the
cluster analysis including the four additional systems with
information on {\Lya} (without regard to whether {\CIV} was measured or
an upper limit)A.  
In this analysis, the {\CIV} upper limits were treated as measured
values.
The class memberships were unchanged for the original 30 systems.
The added systems were classified as Single/Weak based upon their
small $W_{r}({\MgII})$ and narrow kinematics.
This was expected, given the already large spread in $W_{r}({\CIV})$
for the Single/Weak class.

We thus conclude that the clustering results are robust; they are not
affected by biases nor are they sensitive to changes in the sample.

\subsection{Absorption Strengths}

In Figure~\ref{fig:ewall}, we present the rest--frame equivalent
widths (taken from Table~\ref{tab:lineprops} and from Tables 3 and 4
of Paper~I\nocite{paperI}) vs.\ $W_{r}({\MgII})$.
We have also included $W_{r}({\CaII})$ vs.\ $W_{r}({\MgII})$.
The panels are ordered by increasing ionization potential from the
upper left to the lower right and the data point types are the same as
in Figure~\ref{fig:k-means}.

For the non--DLA/{\HI}--Rich systems, note that the absorption
strengths of the low ionization species {\MgI}, {\FeII}, {\SiII},
{\AlII}, and {\CII} increase with increasing $W_{r}({\MgII})$,
indicating these species arise in the phase giving rise to the {\MgII}
absorption.
Overall, the higher ionization species {\SiIV} and {\CIV} exhibit more
scatter, as quantified by the dispersion per observed range. 

We caution that the classification scheme introduced above should
not be taken to suggest that {\MgII} absorbers group into discretized
classes.
Discretization is a byproduct of clustering analysis.
In fact, the distribution functions of the equivalent widths plotted
in Figure~\ref{fig:ewall} are characterized by single modes and
decreasing tails\footnote{The exceptions are $W_{r}({\MgI})$,
$W_{r}({\FeII})$, and $W_{r}({\Lya})$, which are bimodal due to the
DLA/{\HI}--Rich class. However, it is not clear if this bimodality is
due to small numbers and/or is due to the relatively large number of
DLAs in our sample.}
As such, any single absorption property, viewed in this univariate
fashion, is distributed continuously.
However, from the perspective of a multivariate analysis, it is clear
that {\it the overall properties of\/ {\MgII} absorbers group in well
defined regions of a ``multi--dimensional space''}.



\section{Inferring Ionization Conditions and Kinematics}
\label{sec:multiphase}

To examine the kinematic and ionization conditions, we define a new
subsample, IC (IC = Ionization Conditions).
Sample IC includes only those systems with (1) an {\MgII} equivalent
width detection threshold less than or equal to $0.02$~{\AA}, and (2) 
no unresolved saturation in at least four adjacent pixels (1.33
resolution elements) in the {\MgII} $\lambda 2796$ profile.
These selection criteria enforce a high level of accuracy in the
number of Voigt profile components, $N_{cl}$, and their column densities,
velocities, and $b$ parameters.
Simulations of blended, multiple component {\MgII} profiles with these
characteristics show that the distribution of Voigt profile parameters output
from Voigt profile fitting is consistent (99\% confidence) with those used to
generate the profiles (\cite{thesis}).
Data with lower signal--to--noise ratios result in a slight paucity of
Voigt profile components, and therefore (to compensate) the resulting components
have column densities and $b$ parameters that are too large.
Profiles with severely saturated cores (e.g.\ DLA/{\HI}--Rich
absorbers) systematically have fewer Voigt profile components by a factor of
three and have unconstrained column densities and $b$ parameters.
The equivalent width threshold criterion directly translates to a 
signal--to--noise ratio criterion of $S/N=22$ per resolution element
in the continuum.
Sample IC membership is given in column nine of
Table~\ref{tab:lineprops}.

\subsection{Ionization Conditions}

\subsubsection{Single/Weak Systems}

Normally, it is difficult to interpret the ionization conditions
based upon equivalent widths because simple curve of growth arguments
are muddled by the possibility of unresolved saturation in multiple
component absorption.
However, these arguments {\it can\/} hold for the Single/Weak clouds.

As shown by Churchill \etal (1999a\nocite{weak}), under the assumption
of photoionization, Single/Weak clouds with $W_{r}({\MgII}) \leq
0.15$~{\AA}, cannot give rise to $W_{r}({\CIV}) \geq 0.2$~{\AA} (see
their Figure~12).
The upper limit on $W_{r}({\CIV})$ is significantly more restrictive 
when {\FeII} is detected in the {\MgII} cloud (i.e.\ the cloud is
constrained to have low ionization conditions).
Based upon this analysis, we find that roughly half of the observed
Single/Weak absorbers likely have multiple ionization phases, with
{\CIV} and some portion of the {\Lya} absorption arising in spatially
distinct, higher ionization  material.
These findings are confirmed in a thorough study using CLOUDY
(\cite{ferland}) photoionization models tuned to the measured {\MgII}
column densities and constrained by the full complement of ionization
species and transitions covered in FOS spectra (\cite{weakII})

\subsubsection{Multicloud Systems}
\label{sec:cloudymodels}

We now use photoionization modeling to investigate whether multiphase
ionization can also be inferred for the multicloud Classic,
{\CIV}--deficient, and Double systems.
We consider whether the measured {\CIV} and/or {\SiIV} absorption can
or cannot all arise in the {\MgII} clouds, even under extreme
assumptions about their ionization conditions.
The following is based upon the methods employed by Churchill \&
Charlton (1999\nocite{q1206}) in their study of the $z=0.9276$ systems
toward PG~$1206+459$.

We assumed that the {\MgII} clouds in a given system are in
photoionization equilibrium and used CLOUDY (\cite{ferland}) to model
their ionization conditions.
A Haardt \& Madau (1996\nocite{haardtmadau96}) extragalactic spectrum
at $z=1$ was used for the ionizing radiation.
Based upon experiments with various single and extended starburst
galaxy spectral energy distributions (\cite{bruzual}) with assumed
high photon escape fractions, we determined that CLOUDY models
involving {\MgII}, {\FeII}, {\SiIV}, and {\CIV} are not highly
sensitive to the chosen spectrum for $z \sim 1$ (\cite{q1206}); thus
our general conclusions are not sensitive to the assumed spectral
energy distribution.

We obtained a predicted {\CIV} and {\SiIV} equivalent width for each
system by tuning a CLOUDY model to the observed {\MgII} and {\FeII}
column densities in each of its clouds.
We then measured the absorption strengths in synthesized HIRES spectra
using the velocities and $b$ parameters thermally scaled from those
measured for the {\MgII} clouds [see Churchill \& Charlton
(1999\nocite{q1206}) for details.]

Since our goal is to infer if multiphase ionization structure is
possibly present in a given system, we have forced the models to
yield the maximum amount of {\CIV} and {\SiIV} that can arise in the
{\MgII} clouds.  
If a cloud has detected {\FeII}, then, for a given $[\alpha/{\rm Fe}]$
abundance pattern under the assumption of negligible dust depletion, 
the {\FeII} and {\MgII} column densities uniquely
determine the ionization parameter (the logarithm of the ratio of the
number density of hydrogen ionizing photons to the number density of
hydrogen), nearly independent of metallicity.
The ionization parameter uniquely dictates the {\CIV} and {\SiIV}
column densities in the clouds.
Typical ionization parameters in the clouds with {\FeII} are $-4$ to
$-3$. 
We use a solar $[\alpha/{\rm Fe}]$ abundance pattern, since  
this yields a higher ionization parameter for a given ratio of 
{\MgII} to {\FeII} column densities, and thus yields a maximized
{\CIV} and {\SiIV} strength.
If there is no detected {\FeII} for a cloud (either a limit or no
coverage), then the ionization parameter is pushed to as high a value
as possible without the cloud size exceeding 10~kpc.
Typically, this occurs at an ionization parameter of $\sim -1.9$.
A solar metallicity was assumed for all clouds, again in order to
maximize the predicted {\CIV} and {\SiIV} strengths (for a high
metallicity cloud, the hydrogen column density required to give rise
to the observed {\MgII} is smaller, and the ionization parameter can
be pushed to a higher value).
We did not apply any other constraints from the data (e.g.\ {\SiII},
{\CII}, etc.) than those described above.

In Figure~\ref{fig:multi}, we present the predicted maximum equivalent
widths vs.\ the observed equivalent widths for {\CIV} (left panel) and
{\SiIV} (right panel).
The mean error in the measured values is depicted by the open--box
data point in the upper left of each panel.
Diagonal lines demarcate those systems that are inferred to have
multiphase ionization (lower right) from those that could be single
phase.
The maximum predicted {\CIV} strength that can
arise in photoionized {\MgII} clouds is more than $3~\sigma$ short of
the observed $W_{r}({\CIV})$ for a number of systems.
Even two of the {\CIV}--deficient systems are below the line.
This implies that even though these particular systems have below
average {\CIV} absorption strengths, the {\CIV} may still arise in a
distinct, higher ionization phase and not in the {\MgII} clouds.
For {\SiIV}, it appears that fewer systems would require multiphase
structure in this ionization species; {\SiIV} likely resides in the
{\MgII} clouds in a larger fraction of {\MgII} selected absorbers than
does the {\CIV}.

\subsection{{\MgII} and {\CIV} Kinematics}
\label{sec:dataresults}

\subsubsection{Number of Clouds}

In Figure~\ref{fig:nclouds}, we show these same absorption strengths
vs.\ the number of clouds, or subcomponents, $N_{cl}$, obtained from
Voigt profile decomposition.
The data point types are the same as for Figure~\ref{fig:ewall}.

There is a strong correlation of $W_{r}({\MgII})$ with $N_{cl}$, the
number of Voigt profile components (also see \cite{pb90}).
A linear least--squares fit, which we have
drawn as a dotted line in Figure~\ref{fig:nclouds}, yielded a slope of
$0.069\pm0.005$ {\AA}~cloud$^{-1}$.
Note that this relationship is likely resolution dependent.
Also note that the fit is strongly influenced by the two Double
systems with large $N_{cl}$.

We find that there is not as ``linear'' a dependence for
$W_{r}({\FeII})$ nor for $W_{r}({\CIV})$ with {\MgII} kinematics.
There is, however, a tight correlation of $W_{r}({\Lya})$ with the
number of {\MgII} clouds.
We obtained a linear least--squares fit (drawn as a dotted line)
with slope $0.57\pm0.08$ {\AA}~cloud$^{-1}$.
{\it The linear relationship for $W_{r}({\MgII})$ vs.\ $N_{cl}$ suggests
that majority of neutral hydrogen equivalent width arises in the\/
{\MgII} clouds themselves}.

\subsubsection{{\MgII} Kinematics}
\label{sec:kinprops}

In Figure~\ref{fig:kinematics}, the absorption strengths of {\MgII},
{\FeII}, {\Lya}, and {\CIV} are plotted vs.\ the {\MgII} kinematic
spread, $\omega_{v}$.
The data point types are the same as for Figure~\ref{fig:ewall}.

In Figure~\ref{fig:kinematics}, the scatter in $W_{r}({\MgII})$ arises
because the {\MgII} equivalent width is dominated by the largest
clouds, which are usually clustered in subgroups with
velocity spreads less than $\sim 20$~{\kms} [see Figure 12 of Charlton
\& Churchill (1998\nocite{kinematicpaper})].
The kinematic spread, on the other hand, is sensitive to the presence
of small $W_{r}({\MgII})$ clouds with large velocities ($\omega _v
\propto v^{2}$).
The ``limit'' along the upper left of the data, shown as a dotted
line, is due to saturation in the profile line cores; there is a
maximum equivalent width for a given velocity spread when the profile
is black bottomed.
The ``limit'' along the lower right is due to detection sensitivity;
for the signal--to--noise ratios of the sample, there is a minimum
detectable $W_{r}({\MgII})$  for a given velocity spread.
Higher quality data would be required to determine if there is an
actual paucity of systems with large kinematic spreads and extremely
weak {\MgII} absorption.

The strong correlation between $W_{r}({\CIV})$ and $\omega _{v}$ has
previously been discussed by Churchill \etal (1999b\nocite{letter}).
The correlation is driven by the three Double systems
($z=0.8514$ toward Q~$0002+051$, $z=0.9110$ toward PKS~$0823-223$, and 
$z=0.9276$ toward PG~$1206+459$) and the two kinematically extreme
Classic systems ($z=1.3250$ toward PG~$0117+213$ and $z=0.7908$ toward
PKS~$2145+067$), which all have $\omega _{v} \geq 75$~{\kms}.
We have drawn in the maximum--likelihood linear fit to the data, which
has slope $15.4\pm0.2$~m{\AA}~$({\kms})^{-1}$.
More data on Double systems at intermediate redshifts, selected by
$W_{r}({\MgII}) > 0.6$~{\AA}, would be useful for determining how
tight the relationship between $W_{r}({\CIV})$ and {\MgII} kinematics
remains for the largest kinematic spreads.

Three of the five Classic systems above the correlation line are from
the lowest signal--to--noise spectra; there could be missed, small
equivalent width components resulting in a small $\omega_{v}$.
Even so, there is a significant scatter in $W_{r}({\CIV})$ at
a given $\omega _{v}$ for $\omega_{v} \leq 60$~{\kms}.
The Single/Weak systems have $W_{r}({\CIV}) \simeq 0.5$~{\AA} and
$\omega _{v} \simeq 5$~{\kms}, whereas the {\CIV}--deficient systems
have $W_{r}({\CIV}) \leq 0.2$~{\AA} and $\omega _{v} \simeq
45$~{\kms}.

\subsubsection{{\CIV} Kinematics}
\label{sec:c4phase}

In Figure~\ref{fig:broadc4} we display the suite of {\CIV} doublets
feasured in FOS spectra (we exclude ground based data; see Paper~I) that are
not blended with other transitions. 
The velocity window for the {\CIV} doublets is 2000~{\kms}. 
Above each doublet member are ticks giving the velocities of the
{\MgII} Voigt profile components.
The {\MgII} $\lambda 2796$ profiles are also shown with the Voigt profile models
superimposed on the data, which are ordered by increasing kinematic
spread, $\omega_v$.
The velocity window for the {\MgII} data is 600~{\kms}. 

To demonstrate which profiles are resolved, and to quantify the degree
to which they are resolved, we fit the individual {\CIV} members with
single Gaussians while holding the width constant at the value of the
instrumental spread function, $\sigma = 98$~{\kms}.
We have superimposed these {\it unresolved fits\/} on the {\CIVdblt}
profiles.
We note that several of the spectra were obtained prior to the
refurbishing mission in which COSTAR was installed.
For ``large'' aperture aquisition modes ($1${\arcsec} slit), the
instrumental spread function of the pre--COSTAR instrument has extended
wings.
PKS~$0454-220$ is the only object acquired in this mode; thus, the
unresolved fit to the $z=0.4744$ system should be viewed with some
discretion.
Several of the unresolved fits are consistent with the {\CIV} data,
suggesting that the adopted instrumental spread function accurately
represents the pre--COSTAR and post--COSTAR instrument for the
smaller apertures modes.

We computed the quantity, 
\begin{equation}
F_{r} = \frac{W({\rm measured})- W({\rm unresolved})}{W({\rm measured})},
\end{equation}
where $W({\rm measured})$ is the measured {\CIV} equivalent width
adopted for Table~\ref{tab:lineprops}, and $W({\rm unresolved})$
is the {\CIV} equivalent width from the unresolved fits.
Thus, $F_{r}$ measures the fraction of the integrated flux due to
resolved velocity structure in the {\CIV} doublet.
We have noted the $F_{r}$ value and its $1~\sigma$ uncertainty for each
{\CIV} profile in Figure~\ref{fig:broadc4}.
There is a $3~\sigma$ correlation between $F_{r}$ and $\omega_v$.
{\it We have thus demonstrated that the tight $W_{r}({\CIV})$ vs.\
$\omega_v$ correlation is due to resolved velocity structure in the
{\CIV} absorbing gas}.

Velocity structure is not unexpected.
Observed at higher resolution, {\CIV} profiles from Milky Way halo gas
exhibit velocity structure over a $\sim 200$~{\kms} spread, though the
components are typically blended together (\cite{sem00}).
For higher redshift galaxies, the {\CIV} profiles also show distinct
velocity structure (\cite{pb94}) with velocity spreads up to 200~{\kms}.  
In Figure~\ref{fig:q1331c4}, we present an example of a kinematically
complex {\CIV} doublet ($\sim 9$~{\kms} resolution) and two {\FeII}
transitions ({\MgII} was not covered) for the damped {\Lya} absorber
at $z=1.7763$ toward Q$1331+170$\footnote{Interestingly, this DLA has
$W_{r}({\CIV})= 1.63$~{\AA}, twice that of the DLA/{\HI}--Rich systems
in our lower redshift sample. It has $W_{r}({\MgII}) = 1.3$~{\AA}
(Steidel \& Sargent 1992)}.
The ticks above the {\CIV} profiles were obtained by a crude Voigt
profile fit to
the strongest components to parameterize the kinematics; we used this
fit to synthesize the {\CIV} doublet as it would be observed with FOS
resolution and pixel sampling (infinite signal--to--noise ratio).
This is shown in the separate, bottom panel of
Figure~\ref{fig:q1331c4}, with the {\CIV} $\lambda 1548$ and $\lambda
1551$ component ticks above the continuum (not the {\MgII} ticks as in
Figure~\ref{fig:broadc4}). 
An unresolved fit is superimposed on the synthetic FOS doublet for
which we measured $F_{r} = 0.27\pm0.02$.
This value is comparable to $F_{r}$ for the $z=0.5505$ system toward
Q~$1241+174$ and the $z=0.9110$ double system toward PG~$0823-223$.

In cases where the {\MgII} profiles are highly asymmetric, it appears
that the {\CIV} profiles exhibit similar asymmetry in the same sense.
That is, the flux discrepancies between the unresolved fits and the
{\CIV} profiles are aligned in velocity with the kinematic outliers
seen in {\MgII}.
Since we have demonstrated that, in most cases, the {\CIV} does not
arise in the same phase as the {\MgII} clouds (see
\S~\ref{sec:cloudymodels}), {\it we infer that the {\CIV} often arises
in a physically distinct structure from the {\MgII}, but is aligned
kinematically with the {\MgII}}.

\section{Absorption--Galaxy Properties}
\label{sec:galaxies}

We tested for correlations between the galaxy properties and all the
absorption line data presented in Paper~I\nocite{paperI}, including
all possible combinations of equivalent width ratios.
We performed Kendall and Spearman non--parametric rank correlation
tests using the program ASURV (\cite{lavalley}).
The ASURV algorithm provides for measurements that are either upper or
lower limits.

Out of 114 tests performed, seven resulted in correlations
above the $2.5~\sigma$ level\footnote{For 114 correlations tests, only
one test should result in a $2.5~\sigma$ or greater significance level
at random.}; these are:
$W_{r}({\MgII})$, {\MgII} doublet ratio, and 
$W_{r}({\SiIV})/W_{r}({\CIV})$ vs.\ impact parameter;
$W_{r}({\CII})$ and $W_{r}({\MgII})$ vs.\ $M_{K}$; and 
$W_{r}({\Lya})/W_{r}({\CIV})$ and $W_{r}({\SiII})/W_{r}({\SiIII})$
vs.\ $B-K$.
When DLA/{\HI}--Rich systems were removed from the sample, the
significance levels dropped slightly below $2.0~\sigma$.
No test resulted in a correlation at a $3.0~\sigma$ or greater
significance level.

\subsection{Global Ionization and Density Structure?}

In Figure~\ref{fig:galprops} we present plots of the 
absorption properties used in our multivariate analysis, $\omega_{v}$,
$W_{r}({\MgII})$, $W_{r}({\FeII})$, $W_{r}({\Lya})$, and
$W_{r}({\CIV})$, vs.\ host galaxy properties.

$W_{r}({\MgII})$ is anti--correlated with impact parameter (decreases
with galactocentric distance) with a $2.7~\sigma$ significance and
is correlated with $M_{K}$ (increases with decreasing luminosity),
also with a $2.7~\sigma$ significance.
Both these trends arise because DLAs, which give rise to the largest 
$W_{r}({\MgII})$, are often observed at low impact parameters and
often have low luminosities ($M_{K} > -24$).

The $W_{r}({\CIV})$ vs.\ $K$ luminosity relationship can provide a
test for halo models in which a virialized hot phase ($T\sim
10^{6}$~K) pressure confines the low ionization phase (e.g.\
\cite{mo96}).
These models predict an anticorrelation between $W_{r}({\CIV})$ and
galaxy mass (i.e.\ $K$ luminosity); {\CIV} is predicted to be enhanced
in smaller, lower mass galaxies, which have low gas pressures in their
halos.
As seen in Figure~\ref{fig:galprops}, there is a visual, yet not
statistically significant, trend for the most massive galaxies to have
smaller $W_{r}({\CIV})$ in our small sample.
This is driven by the high luminosity [and very red; see $W_{r}({\CIV})$
vs.\ $B-K$] {\CIV}--deficient systems.

In Figure~\ref{fig:galimpacts}, we have plotted the equivalent width
ratios of {\SiIV} to {\CIV}, {\Lya} to {\CIV}, {\Lya} to {\MgII},
{\FeII} to {\CIV}, and the {\CIV} and {\MgII} doublet ratios vs.\
impact parameter.
We have selected combinations only of the gas properties used in our
multivariate analysis, with the exception of the ratio
$W_{r}({\SiIV})/W_{r}({\CIV})$.
This ratio is anticorrelated with impact parameter at the $2.8~\sigma$
level when all galaxies are included and at the $2.1~\sigma$ level when
DLA/{\HI}--Rich systems are excluded.
If this trend were to hold for a larger sample, it would be consistent
with the scenario in which halo gas is photoionized by the
extragalactic UV background flux resulting in a increasing ionization
level with increasing galactocentric distance (see \cite{savage97}).

The {\MgII} doublet ratio is correlated with impact parameter at the
$2.7~\sigma$ significance level.
This reflects the fact that the gas tends to become optically thin at
larger galactocentric distances, where $W_{r}({\MgII})$ is
systematically smaller.
Again, the DLAs dominate the trend.

The ratio $W_{r}({\Lya})/W_{r}({\MgII})$ is not correlated with impact
parameter when all galaxies in the sample are included.
However, when the DLA/{\HI}--Rich systems are excluded a $2.5~\sigma$
level correlation is observed.
This is governed by the general decrease in {\MgII} equivalent widths
with impact parameter, since {\Lya} shows no trend with impact
parameter (when DLA/{\HI}--Rich systems are excluded).
This particular trend is difficult to interpret because of the very
different curve of growth behavior for {\MgII} and {\Lya}.
It is not possible to address the possibility of a metallicity or
ionization gradient.
The location of the Lyman limit break was covered for only six of the
galaxies and was detected for all six, sampling impact parameters
from $10$ to $40$~kpc.  

\subsection{Ionization Conditions and Colors}

In Figure~\ref{fig:galcolors}, we have plotted the {\CIV} to {\MgII}, 
{\Lya} to {\CIV}, and {\Lya} to {\MgII} equivalent width ratios vs.\
host galaxy $B-K$ color.
Both $W_{r}({\Lya})/W_{r}({\CIV})$ and $W_{r}({\Lya})/W_{r}({\MgII})$
are correlated with $B-K$ at a $2.6~\sigma$ significance level.
This trend in our small sample is governed by both the
DLA/{\HI}--Rich systems and the {\CIV}--deficient systems being
associated with redder galaxies.

\subsection{Correlation Test: Overall Results}

The degree of scatter in the absorption properties presented here are 
similar to that found by Churchill \etal (1996\nocite{csv96}) for the
{\MgII} absorption properties only.
None of the presented absorption properties correlated with the galaxy
properties at a {\it high\/} significance level.
This is not to say that definite trends, or even statistically
significant correlations, do not exist.
The large scatter in the properties signifies large local variations
from line of sight to line of sight through the galaxies; however,
even with our small sample, we have uncovered some suggestive global
trends.

We also find that the inclusion or exclusion of DLA/{\HI}--Rich systems
in any given test significantly alters the statistics.
The host galaxies of DLAs are seen to have a wide range of
luminosities, morphologies, colors, and surface brightnesses (e.g.\ 
\cite{lebrun97}; \cite{rao98}), and do not represent a homogenous
population, despite the fairly homogeneous absorption properties by
which they are selected.

\section{Discussion}
\label{sec:discussion}

We reemphasize that a taxonomic scheme based upon a multivariate
clustering analysis will, by its very nature, result in a
discretization of what may actually be a continuum of properties.
Nonetheless, if we adopt the clustering analysis results at face value
(see \S~\ref{sec:clusteringdiscussion}) and segregate the wide variety of
{\MgII} absorber properties into individual ``classes'', we may be
able to find order in an otherwise complex array of gas properties and
kinematics.
The hope is to gain further insights into defining distinctive or
``characteristic'' properties of {\MgII} absorbers and to
understanding the galactic processes that give rise to the observed
range of these ``characteristics''.

What follows is primarily a speculative discussion, given the fact that
there is still very little data upon which inferrences can be based.

\subsection{Classic vs.\ {\CIV}--deficient Systems} 

The existence of {\CIV}--deficient systems with $W_{r}({\CIV})\leq
0.15$~{\AA}, which have kinematics similar to the Classics, indicates
that the connection between {\CIV} absorption and {\MgII} kinematics
does not operate the same in all galaxies (or at all locations in
galaxies).

There are only three {\CIV}--deficient systems with measured galaxy
properties ($z=0.4297$ toward PKS~$2128-123$, $z=0.6601$ toward
Q~$1317+277$, and $z=0.7291$ toward PG~$0117+213$).
The host galaxies of these three systems are reddish ($B-K = 3.3$,
$3.8$, and $4.0$) and are among the more luminous ($M_K = -24.7$,
$-25.7$, and $-26.3$).
Also, they are probed at large impact parameters ($D = 32$, $38$, and
$36$ $h^{-1}$~kpc), respectively (see Figure~\ref{fig:galprops}).
The redder colors and large luminosities would suggest massive,
early--type galaxies.
Based upon {\it HST} images of the $z=0.6601$ and $z=0.7291$ absorbers
(\cite{steidel98}), we roughly (visually) classify them to be an
edge--on S0 galaxy (probed on its minor axis) and a face--on SBa
galaxy, respectively.

Despite small numbers, these relatively high masses, early--type
morphologies, and large impact parameters may be a clue to the
observed spread in $W_{r}({\CIV})$ for a given {\MgII} absorption
profile type.
Semianalytical models of pressure confined gaseous halos predict
smaller {\CIV} strengths in more massive galaxies (e.g.\ \cite{mo96}),
and such a trend is not inconsistent with our data.
The weak {\CIV} cannot simply be due to the line of sight through 
the galaxy, because some Classics at large impact parameters have very
strong {\CIV} strengths.
Also, based upon cross--sectional arguments from the number of {\CIV}
absorbers per unit redshift (\cite{steidel93}) and upon the expected
lower pressures and gas densities at larger galactocentric distances
(\cite{mo96}), it is expected that {\CIV} absorption would be strong
at large impact parameters.

Perhaps strong {\CIV} is expected in regions of relatively pronounced
star formation, which would be consistent with {\CIV}--deficient
absorbers being associated with earlier--type galaxies.
Alternatively, in {\CIV}--deficient absorbers, the {\CIV} could be
ionized away due to collisional processes, such as with the intragroup
medium scenario proposed by Mulchaey \etal (1996\nocite {mulchaey}).
If so, strong {\OVI} absorption would be predicted; unfortunately, we
cannot address {\OVI} absorption for the {\CIV}--deficient systems
[only one is ``measured'' to have an unrestrictive upper limit on
$W_{r}({\OVI})$].
A further possibility is that galaxies with extended X--ray emission,
usually early--type galaxies (e.g. \cite{mathews98}), may
preferentially be {\CIV}-- and {\OVI}--deficient because of the
extremely hot environment.
If so, the {\MgII} absorbing gas would necessarily arise in {\HI} gas,
which is of external origin for ellipticals from galaxy mergings
(\cite{knapp85}; \cite{vangorkom92}) and is sometimes found to be in
disk--like structures extending 5 to 10 times the optical radii in
``dust--lane'' elliptical galaxies (e.g.\ \cite{morganti})
Thus, one might hypothesize that {\MgII} absorption associated with
elliptical galaxies arises preferentially in post--merger ellipticals.

\subsection{Classic vs.\ Double Systems} 

There are at least two obvious explanations for the Double systems;
they might be two Classic systems clustered in line--of--sight
velocity (bound or unbound), or a primary galaxy and a satellite
(e.g.\ possible an interacting LMC--like object).

The three Double systems in our sample have the largest
$W_{r}({\CIV})$, which are consistent with that of two typical Classic
systems. 
Alternatively, they could arise in a primary galaxy either undergoing
a minor merger or residing in a group with several minor galaxies.
Using the Local Group as a model and applying the simple
cross--sectional dependence for $W_{r}({\MgII})$ with galaxy luminosity
(\cite{steidel95}; also see \cite{mclin98}), we estimated the
probability of intercepting a ``double'' absorber for a random line of
sight passing through the Milky Way (line--of--sight kinematics were
not considered).
We found a $25$\% chance of intercepting both the LMC and the Milky
Way, and a $5$\% chance of intercepting both the SMC and the Milky Way.  
All other galaxies in the Local Group have negligible probabilities of
being intercepted for a line of sight passing within 50 kpc of the Milky
Way.

An alternative scenario is that Double systems are otherwise typical
Classic systems, but for the interception of kinematic outlier clouds,
either due to infalling or ejected ``high velocity'' gas.
This picture would naturally invoke elevated star formation to explain
the large $W_{r}({\CIV})$ in Doubles and, possibly, the correlation
between $W_{r}({\CIV})$ and {\MgII} kinematics.
Infalling material can enhance star formation
(\cite{hummel90}; \cite{lutz92}; \cite{hernquist95};
\cite{hibbard96}), which can provide the energetics to support a
multiphase halo (\cite{dahlem98}) with strong {\CIV} absorption [i.e.,
as with the Galaxy (Savage \etal 1997\nocite{savage97})].
Consistent with this scenario would be that the bluest galaxies would
preferentially be associated with Double systems. 
Only one Double system has measured galaxy properties ($z=0.8519$
toward Q~$0002+051$), and we note that its color ($B-K = 2.9)$ is
among the bluest in the sample, consistent with a star forming object.
An {\it HST\/} image (\cite{steidel98}) reveals the galaxy to be
compact with a spherical, featureless morphology (no evidence for
merging).

\subsection{Classic vs.\ DLA/{\HI}--Rich} 

Though the Classic {\MgII} systems are well understood to select
normal, bright galaxies (\cite{sdp94}), the DLA/ systems are seen to
select [based upon $N({\HI}) \geq 2 \times 10^{20}$~{\cmsq}] an
eclectic population of low luminosity, low surface brightness, and
dwarf galaxies (\cite{lebrun97}; \cite{rao98}, 1999\nocite{rao99}).
This might suggest that not all classes of {\MgII} absorber have
a systematic connection with galaxy type or galaxy properties.

What is interesting, however, is that the {\MgII} kinematic spreads
for DLA/{\HI}--Rich systems are tightly clustered around $\sim
60$~{\kms}, have no weak, kinematic outlier clouds, and have
intermediate to weak {\CIV} absorption strengths.
This is in contrast to those at redshifts $2$--$3$, where {\it
strongly\/} absorbing, higher velocity outlying clouds and often
``double'' profiles are characteristic of low ionization DLA profiles
(\cite{lu_dla96}).

Whatever physical process or environment gives rise to kinematic
outlying {\MgII} clouds and strong {\CIV}, it is apparently not
acting in intermediate redshift DLA/{\HI}--Rich systems.
Alternatively, these trends could be due to selection effects of
sightlines through {\HI}--rich environments, which implies a
line--of--sight dependency for absorber class.

\subsection{Absorber Class Evolution?}
 
We have shown that there is a signicant correlation between
$W_{r}({\CIV})$ and the {\MgII} kinematic spread.
But, exactly how do the {\MgII} kinematics vary with $W_{r}({\CIV})$
for a given $W_{r}({\MgII})$?
In Figure~\ref{fig:postage}, we have plotted the {\MgII} $\lambda
2796$ profiles from our HIRES/Keck spectra in the approximate
locations they occupy in the $W_{r}({\CIV})$--$W_{r}({\MgII})$ plane.
Based upon this schematic, we have an impression of how the kinematic
``complexity'' of the absorption profile relates to both the {\MgII}
and {\CIV} absorption strengths.
Roughly, each of the five classes obtained from our multivariate
analysis falls in a well defined region, which we have marked
to guide the reader.
Note that, in general, the {\CIV} strength is largest when the
kinematic ``complexity'' is the greatest; {\CIV} absorption appears to
have no direct connection with the optical depth of the strongest
{\MgII} component, nor with the total number of Voigt profile components.
Most remarkably, it would appear that the {\it gross\/}
characteristics of the {\MgII} kinematics can be predicted simply
based upon the absorber's location on the
$W_{r}({\CIV})$--$W_{r}({\MgII})$ plane.

Since the distribution of {\MgII} equivalent widths evolves with
redshift (\cite{ss92}), and this must be reflected in the {\MgII}
kinematics as well, it seems reasonable to assume that evolution
should be discernable on the $W_{r}({\CIV})$--$W_{r}({\MgII})$ plane.

There are at least two possible types of evolution of absorber
classes:
(1) the number density per unit redshift of a given class could
evolve, either diminishing with or increasing with time, and/or 
(2) any absorber in a given class could evolve into another class.
The former is systematic and would indicate a global, cosmic
evolution, which might imply the existence of absorber classes not
seen in our intermediate redshift sample.
The latter form of evolution would be stochastically related to the
processes occuring in galaxies and their environs.

We present the observational $W_{r}({\CIV})$--$W_{r}({\MgII})$ plane
in Figure~\ref{fig:civmgii}.
The intermediate redshift data (taken from this study) have small, filled
circle data points.
We have included lower and higher redshifts data from the literature,
which we have listed in Table~\ref{tab:civ-literature}.
The higher redshift absorbers ($1.2 \leq z \leq 2.1$) are taken from
Bergeron \& Boiss\'{e} (1984\nocite{bb84}), Boiss\'{e} \& Bergeron
(1985\nocite{boisse85}), Lanzetta \etal (1987), and Steidel \& Sargent
(1992\nocite{ss92}) [six--pointed stars].
The lower redshift points  ($0.1 \leq z \leq 0.6$) are taken from
Bergeron \etal (1994) [downward pointing, open triangles], and the
$z\simeq 0$ data\footnote{We note that the $z\simeq0$ sample is small
and is selected based upon a bright, background quasar having a small
projected separation from a nearby galaxy, as opposed to being
{\MgII}--absorption selected (Bowen \etal 1995); this may have
introduced a bias, possibly related to impact parameter distribution,
as compared to the higher redshift data.} are taken from Bowen \etal
(1995\nocite{bowen95}, 1996\nocite{bowenm61}), Jannuzi \etal
(1998\nocite{cat3}), and Bowen (1999\nocite{bowen99}, private
communication) [open diamonds].
There are four extraordinary systems listed in
Table~\ref{tab:civ-literature} that have $W_{r}({\MgII})$ ranging from
$\sim 4$ to $\sim 7$~{\AA}.
These have not been presented in Figure~\ref{fig:civmgii}.

Three points from Figure~\ref{fig:civmgii} are that
(1) there are absorbers in the higher, lower, and $z\simeq0$ redshift 
samples with {\CIV} and {\MgII} strengths typical of each of the five
classes found in our clustering analysis of the intermediate redshift
systems; 
(2) there are both higher and $z\simeq0$ redshift absorbers occupying
regions of the $W_{r}({\CIV})$--$W_{r}({\MgII})$ that are not
represented in our intermediate redshift sample; and
(3) as with the intermediate redshift systems, the higher redshift
systems exhibit a large  range of $W_{r}({\CIV})$ for a given
$W_{r}({\MgII})$. 
Though there is no suggestion for such a spread in the lower redshift
systems, there are too few measurements to characterize the spread in
$W_{r}({\CIV})$ values.

Given the fairly systematic dependence of kinematics in the
$W_{r}({\CIV})$--$W_{r}({\MgII})$ plane, we can infer that the
{\MgII} kinematics of the lower and higher redshift systems that
occupy locations consistent with a given absorber class are likely to
be similar to that class' kinematics.

The higher redshift systems with $W_{r}({\MgII}) > 2$~{\AA}, are a
class of absorber not present in our sample.
Note the large spread in $W_{r}({\CIV})$ for these systems, which
would suggest that some are ``{\CIV} deficient''.
This may be indicative that the physical processes giving rise to the
range of $W_{r}({\CIV})$ in intermediate redshift systems are also
occuring in these $W_{r}({\MgII}) > 2$~{\AA}, higher redshift systems.

In Figure~\ref{fig:highzMgII}, we present the HIRES/Keck {\MgII}
$\lambda 2796$ profiles of four $z \geq 1$ {\MgII} systems with
$W_{r}({\MgII}) > 2.0$~{\AA}. 
Note that the profiles are optically thick with virtually unity
doublet ratio over a the full velocity span, which is typically
$250~{\kms}$ or greater.
We also show the $z\simeq 0$ {\MgII} profile measured in the spectrum
of SN 1993J in M81 using the GHRS on {\it HST} (\cite{bowen95};
spectrum provided courtesy of Dr.\ D. Bowen).
In comparison to the smaller equivalent width, intermediate redshift
absorbers, the {\MgII} kinematics and absorption strengths of these
$W_{r}({\MgII}) > 2$~{\AA} systems are clearly unique, exhibiting
``double'' black--bottomed profiles.

Of the systems shown in Figure~\ref{fig:highzMgII}, only the
$z=1.7945$ system toward B2~$1225+317$ is represented on
Figure~\ref{fig:civmgii}.
Though the {\CIV} strengths of the other systems shown in
Figure~\ref{fig:highzMgII} are unmeasured or unpublished, it is clear
that these systems constitute their own unique ``class'' (with respect
to the five classes found in our clustering analysis).
As shown in Figure~\ref{fig:highzMgII}, the {\MgII} kinematics are
suggestive of ``double--DLAs'', and, in fact, the $z=1.7945$ system
toward B2~$1225+317$ is a DLA (\cite{bechtold87}).
In high quality data, ``Double--DLAs'' might be expected to have
slightly asymmetric damping wings, reflecting the different column
densities in the two systems.

Steidel \& Sargent (1992\nocite{ss92}) found that the number density
of these large equivalent systems decreases with redshift over the 
range $0.3 \leq z \leq 2.2$ (the full range over which the
{\MgII} doublet is observable from the ground).
Based upon the SN 1993J spectrum of M81 and the Galaxy in absorption
(the line of sight passes through half the disk and halo of M81, half
the disk and halo of the Galaxy, and through ``intergalactic''
material apparently from the strong dwarf--galaxy interactions taking
place with both galaxies), one possibility is that these systems arise
from two galaxies with low impact parameters that happen to have a
line--of--sight superposition.

For this scenario to be consistent with the {\MgII} absorber evolution, 
the number of galaxy pairs would need to decrease in step with the
evolution of the absorbers themselves over the same redshift regime.
In fact, over the redshift interval $1\leq z \leq 2$, it is seen that
the galaxy pair fraction, where pairs are defined to have projected
separations less than $20$~kpc, evolves proportional to $(1+z)^{p}$,
with $2 \leq  p \leq 4$ (\cite{neusch97}, and references therein).
Le~Fevre \etal (1999\nocite{lefevre}) found a similar, better
constrained result with $p = 2.7\pm0.6$ for $0 \leq z \leq 1$.
These compare well with $p = 2.2\pm0.7$ for {\MgII} absorbers with
$W_{r} > 1.0$~{\AA} (\cite{ss92}; the evolution is probably
stronger for those with   $W_{r} > 2.0$~{\AA}).

As such, galaxy pair evolution remains a plausible scenario as a
contribution to the evolution of large equivalent width, higher
redshift {\MgII} absorbers and the class we loosely have called
``Double--DLAs''. 
We note, however, that outflows from very luminous, star bursting
galaxies at these high redshifts could also account for some of these
``super'' systems, since the low ionization gas can be very
prominent in absorption (\cite{pettini99}).


\section{Summary}
\label{sec:summary}
 
We have performed a multivariate analysis of the absorption strengths
and kinematics of {\MgII} and the absorption strengths of {\FeII},
{\Lya}, and {\CIV} for a sample of 45 {\MgII} absorption--selected 
systems.
Descriptions of the survey and analysis of the data used herein have
been presented in Churchill \etal (1997\nocite{thesis},
1999a\nocite{weak}) and in Paper~I\nocite{paperI}.
The multivariate analysis was performed using the STATISTICA package
({\it www.statsoft.com\/}).
We applied both a ``Tree Clustering'' and ``$K$--means Clustering''
analysis to the 30 systems in Sample CA from the data listed in
Table~\ref{tab:lineprops}.

We have also compared the low, intermediate, and high ionization
absorption properties of 16 systems with the $B$ and $K$ luminosities,
$B-K$ colors, and impact parameters of their host galaxies.
The full complement of absorption line data was taken from
Paper~I\nocite{paperI}.
The galaxy properties, listed in Table~\ref{tab:galprops}, are taken
from the survey of Steidel \etal (1994\nocite{sdp94}) and the previous
studies of Churchill \etal (1996\nocite{csv96}) and Steidel \etal
(1997\nocite{3c336}).
We tested for absorption--galaxy property correlations using Kendall
and Spearman non--parametric rank correlation indicators as
implemented with the program ASURV (\cite{lavalley}), which
incorporates upper limits on the data.

\begin{figure*}[t]
\plotone{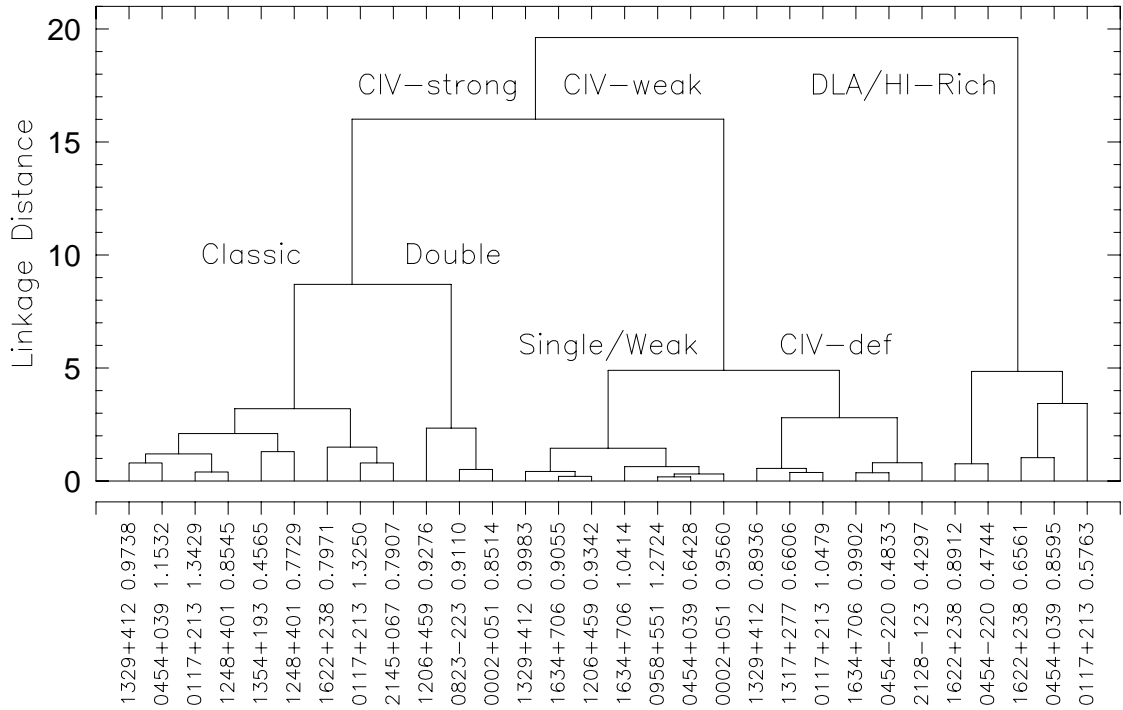}
\caption[fig2.ps]
{Tree cluster diagram showing linkage distance in a
subsample of 30 absorbers, with each labeled across the bottom.  The
discriminating absorption properties are $W_{r}({\MgII})$,
$W_{r}({\Lya})$, $W_{r}({\CIV})$, $W_{r}({\FeII})$, and $\omega_v$,
using a $N(0,1)$ standardization (see text).  A Ward's amalgamation
algorithm was used with Euclidean distance.  The isolated branches
with linkages greater than $\sim 4$ are given classifications based upon 
the cluster means (see Figure~\ref{fig:k-means}).
\label{fig:treecluster}}
\end{figure*}

\begin{figure*}[t]
\plotone{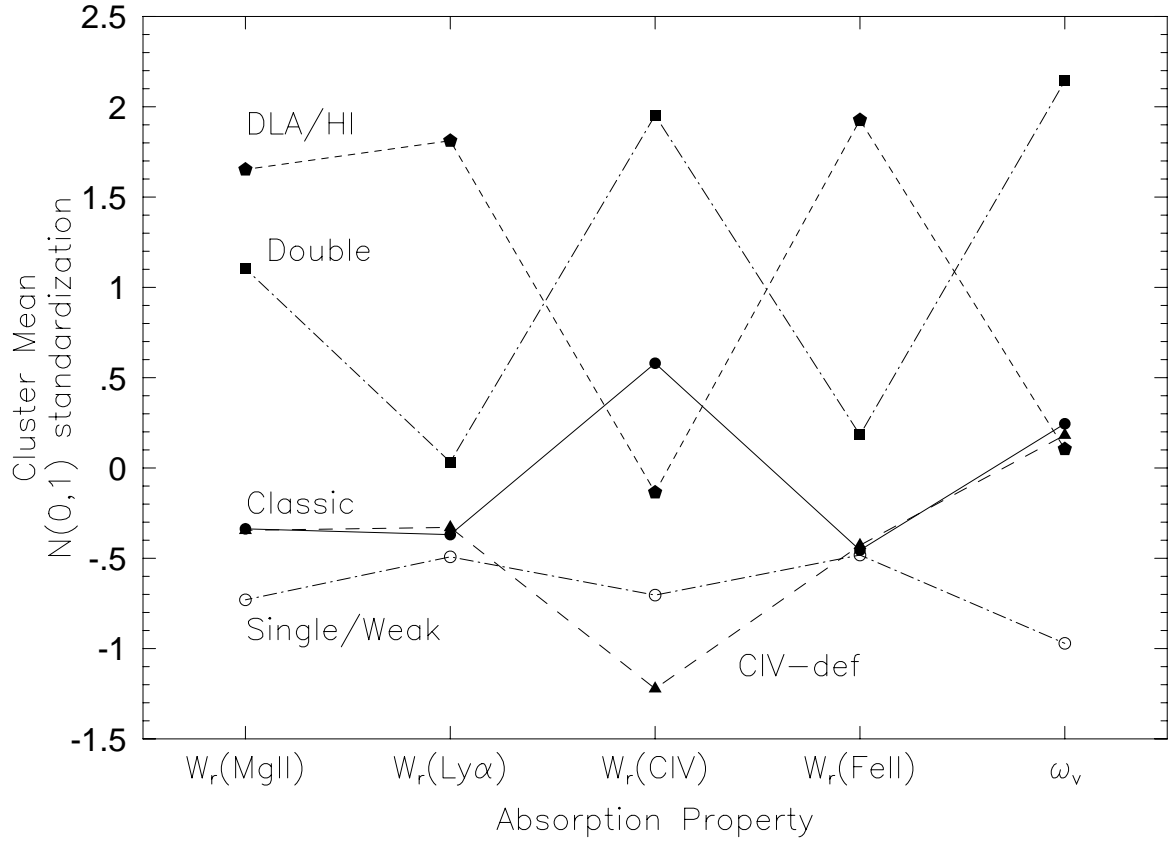}
\caption[fig3.ps]
{$K$--means cluster diagram for $K=5$ showing the mean
values $W_{r}({\MgII})$, $W_{r}({\Lya})$, $W_{r}({\CIV})$,
$W_{r}({\FeII})$, and $\omega_v$, placed on a $N(0,1)$ standardization
(see text), for {\it each\/} of the five clusters shown in
Figure~\ref{fig:treecluster}.
With this standardization, zero is the mean value and $\pm1$ is the
standard deviation for a given absorption property.
The cluster classifications are given on the left hand side.
\label{fig:k-means}}
\end{figure*}

\begin{figure*}[t]
\plotone{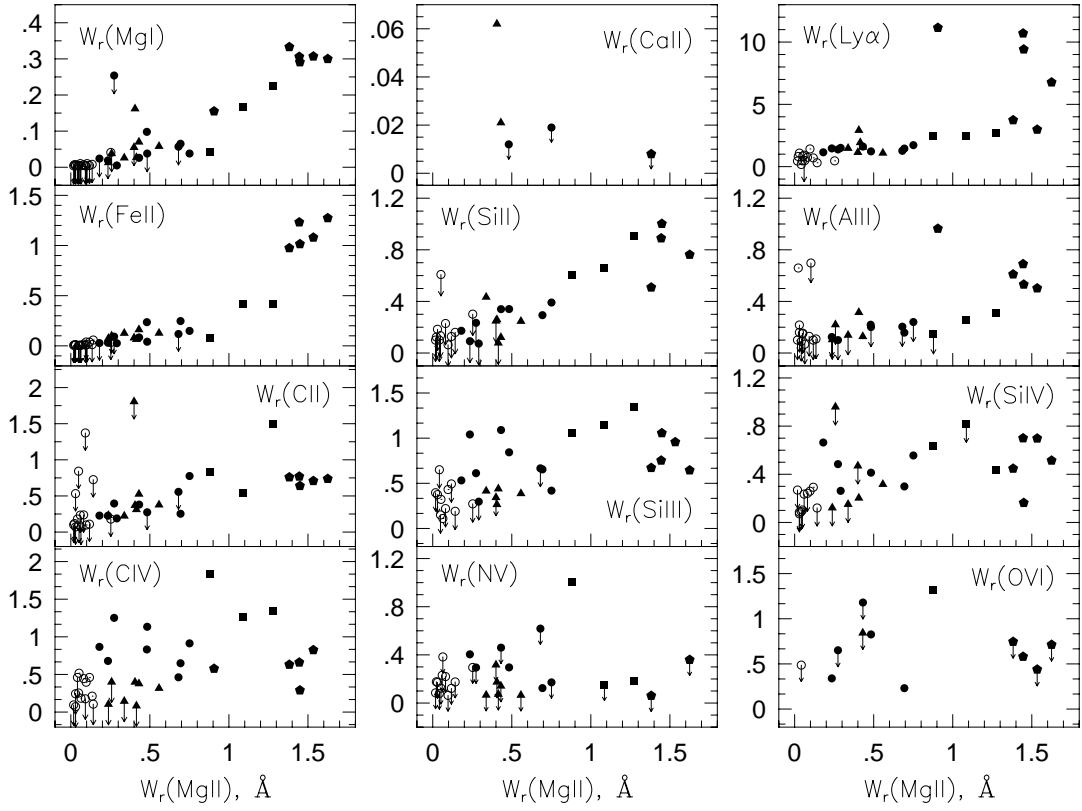}
\caption[fig4.ps]
{The rest--frame equivalent widths of the elemental
species listed in Table~\ref{tab:lineprops} and in Tables 3 and 4 from
Paper I vs.\ that of {\MgII} $\lambda 2796$.
Also included is {\CaII}.
The panels are in order of increasing ionization potential from the
upper left to the lower right.
Data point types denote the various absorber classes, and are the same
as in Figure~\ref{fig:k-means}.  Downward pointing arrows denote upper
limits.
\label{fig:ewall}}
\end{figure*}

\begin{figure*}[t]
\plotone{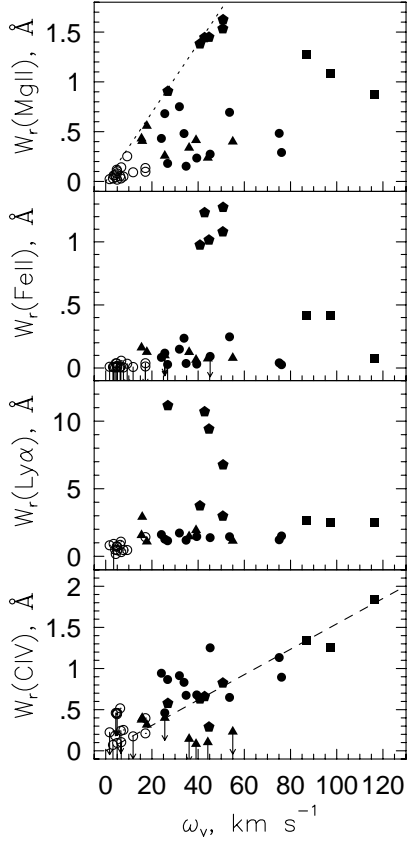}
\caption[fig5.ps]
{The rest--frame equivalent widths of {\MgII},
{\FeII}, {\Lya}, and {\CIV} vs. the {\MgII} kinematic spread, $\omega
_{v}$. Data point types denote the various absorber classes, and are
the same as in Figure~\ref{fig:k-means}.
\label{fig:kinematics}}
\end{figure*}

\begin{figure*}[t]
\plotone{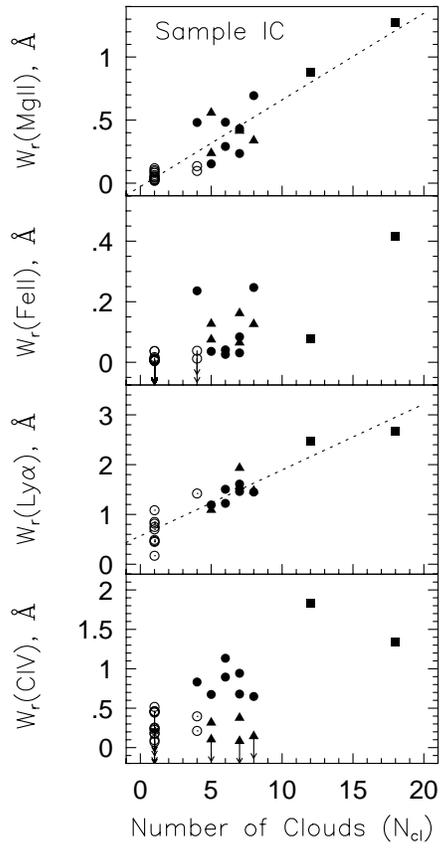}
\caption[fig6.ps]
{The rest--frame equivalent widths of {\MgII},
{\FeII}, {\Lya}, and {\CIV} vs. the number of Voigt profile components, or
clouds, $N_{cl}$.  Only Sample IC is presented for reasons given in
the text.  Data point types denote the various absorber classes, and
are the same as in Figure~\ref{fig:k-means}.
\label{fig:nclouds}}
\end{figure*}

\begin{figure*}[t]
\plotone{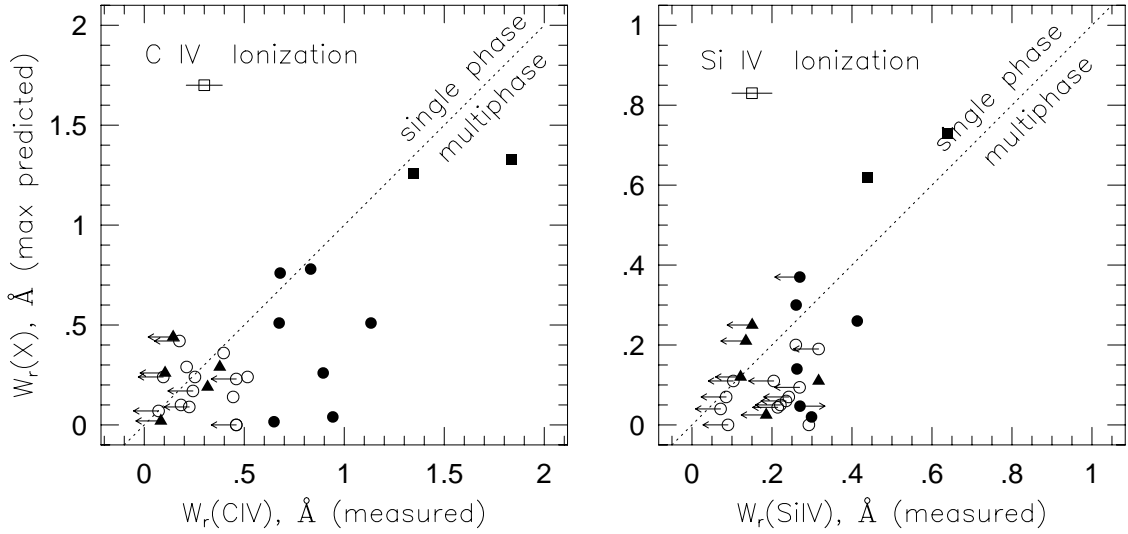}
\caption[fig7.ps]
{The {\it maximum\/} $W_{r}({\CIV})$ (left panel) and
$W_{r}({\SiIV})$, under the assumption of extremely high
photoionization equilibrium, that can arise in single phase {\MgII}
clouds vs.\ the measured rest--frame equivalent widths.
The predicted points are based upon CLOUDY models of the systems
(Sample IC), incorporating the number of clouds and their column
densities, $b$ parameters, and kinematics (see text for details).
If a system lies below the line, then the measured {\CIV} or {\SiIV}
cannot arise solely in the {\MgII} clouds, but must arise in a distinct
highly ionized phase that does not give rise to detectable amounts of
{\MgII}.  The typical error bar of the measure data are given in the
upper left corners of both panels.  Data point types denote the
various absorber classes.
\label{fig:multi}}
\end{figure*}

\begin{figure*}[t]
\plotone{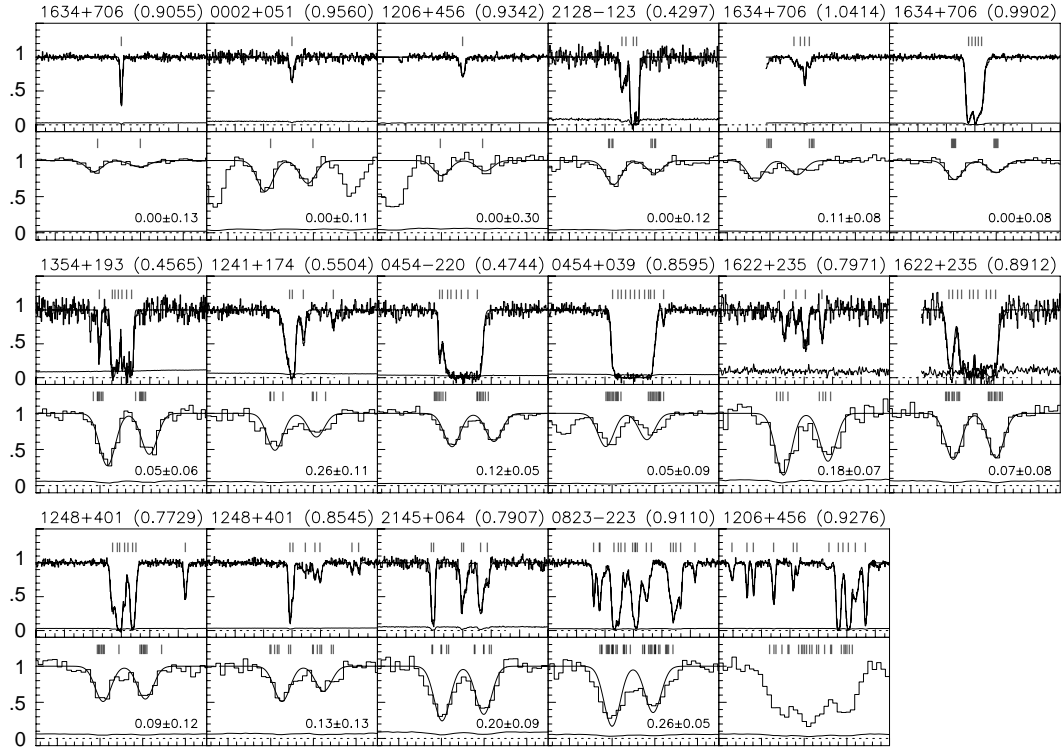}
\caption[fig8.ps]
{The {\CIV} kinematics.  Shown are the {\MgII}
$\lambda 2796$ transitions and {\CIVdblt} doublets of the 17 systems
for which both members of the {\CIV} doublet are detected and
unblended with other transitions in the FOS spectra.
The panels with {\MgII} show a velocity window of $600$~{\kms},
centered on the system zero point.  Ticks above the continuum give the
Voigt profile component velocities and the smooth curves through the data are
the Voigt profile fit results.  The panels with the {\CIV} doublets show a
velocity window of 2000~{\kms}.  The {\MgII} Voigt profile velocities are
projected over the continuum of each doublet.  The smooth curves
through the data are single Gaussian fits with the widths held at the
instrumental resolution, but with the centroids and depths allowed to
vary. 
\label{fig:broadc4}}
\end{figure*}

\begin{figure*}[t]
\plotone{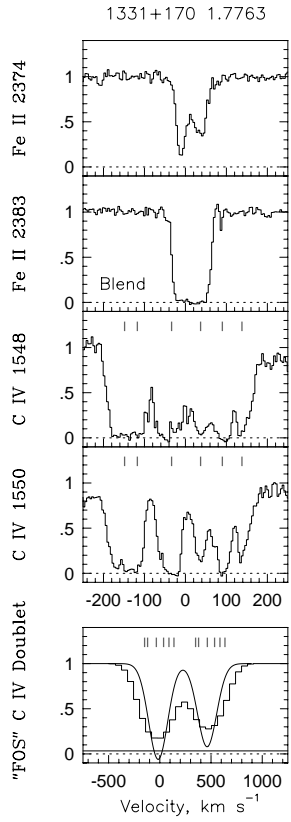}
\caption[fig9.ps]
{The {\FeII} and {\CIV} kinematics of a DLA at
$z=1.7763$ toward Q $1331+170$; the {\MgII} profiles were not covered
due to the setting of the HIRES echelle.  The upper panels show
HIRES/Keck data at $R=22,500$ over a velocity window of 500~{\kms}.
Ticks above the {\CIV} roughly give the kinematic composition of the
high ionization gas.  The lower panel shows simulated FOS spectra of
the {\CIV} doublet over a velocity window of 2000~{\kms}.  As with
Figure~\ref{fig:broadc4}, single Gaussian fits with widths held at the
instrumental resolution are superimposed on the data. \label{fig:q1331c4}}
\end{figure*}

\begin{figure*}[t]
\plotone{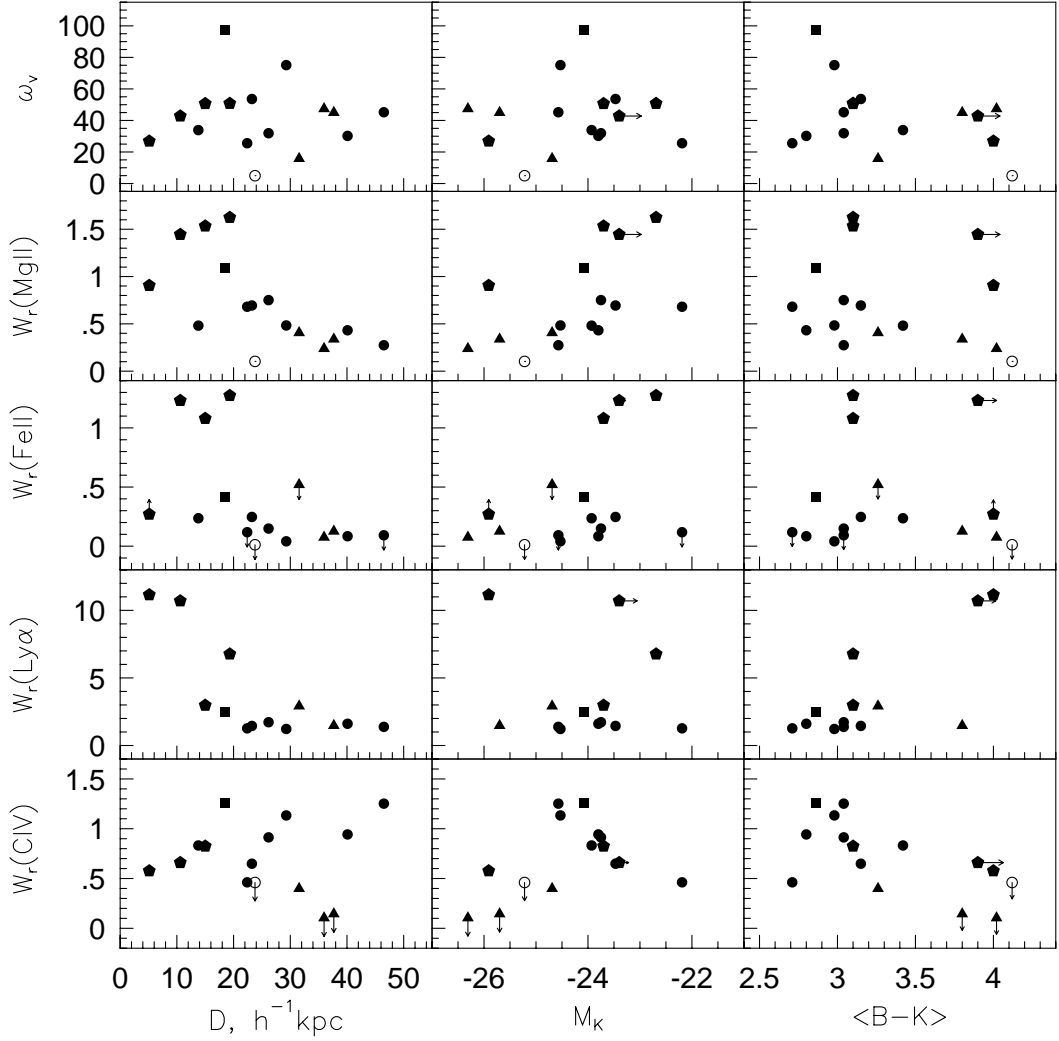}
\caption[fig10.ps]
{The {\MgII} kinematic spread, and {\MgII}, {\FeII},
{\Lya}, and {\CIV} absorption strengths vs.\ host galaxy impact
parameter, absolute $K$ magnitude, and rest--frame $B-K$ color.
\label{fig:galprops}}
\end{figure*}

\begin{figure*}[t]
\plotone{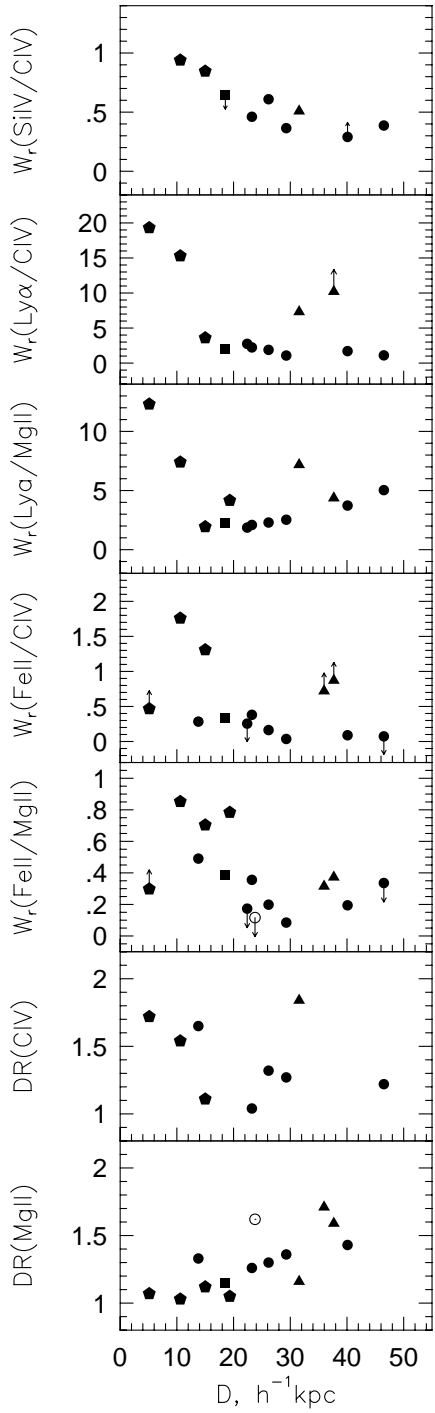}
\caption[fig11.ps]
{Selected equivalent width ratios vs.\ host galaxy
impact parameters.
\label{fig:galimpacts}}
\end{figure*}

\begin{figure*}[t]
\plotone{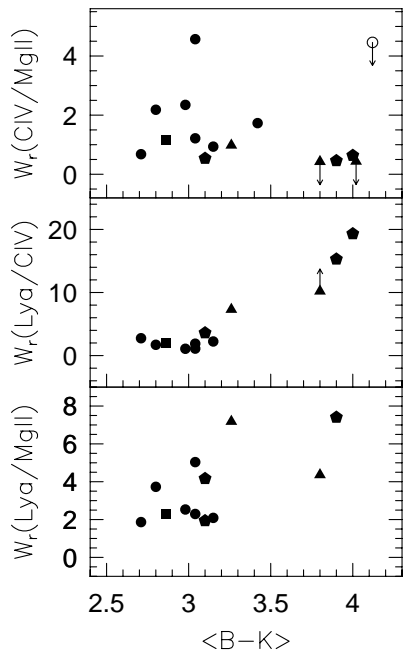}
\caption[fig12.ps]
{Selected equivalent width ratios vs.\ host galaxy
$B-K$ colors.
\label{fig:galcolors}}
\end{figure*}

\begin{figure*}[t]
\plotone{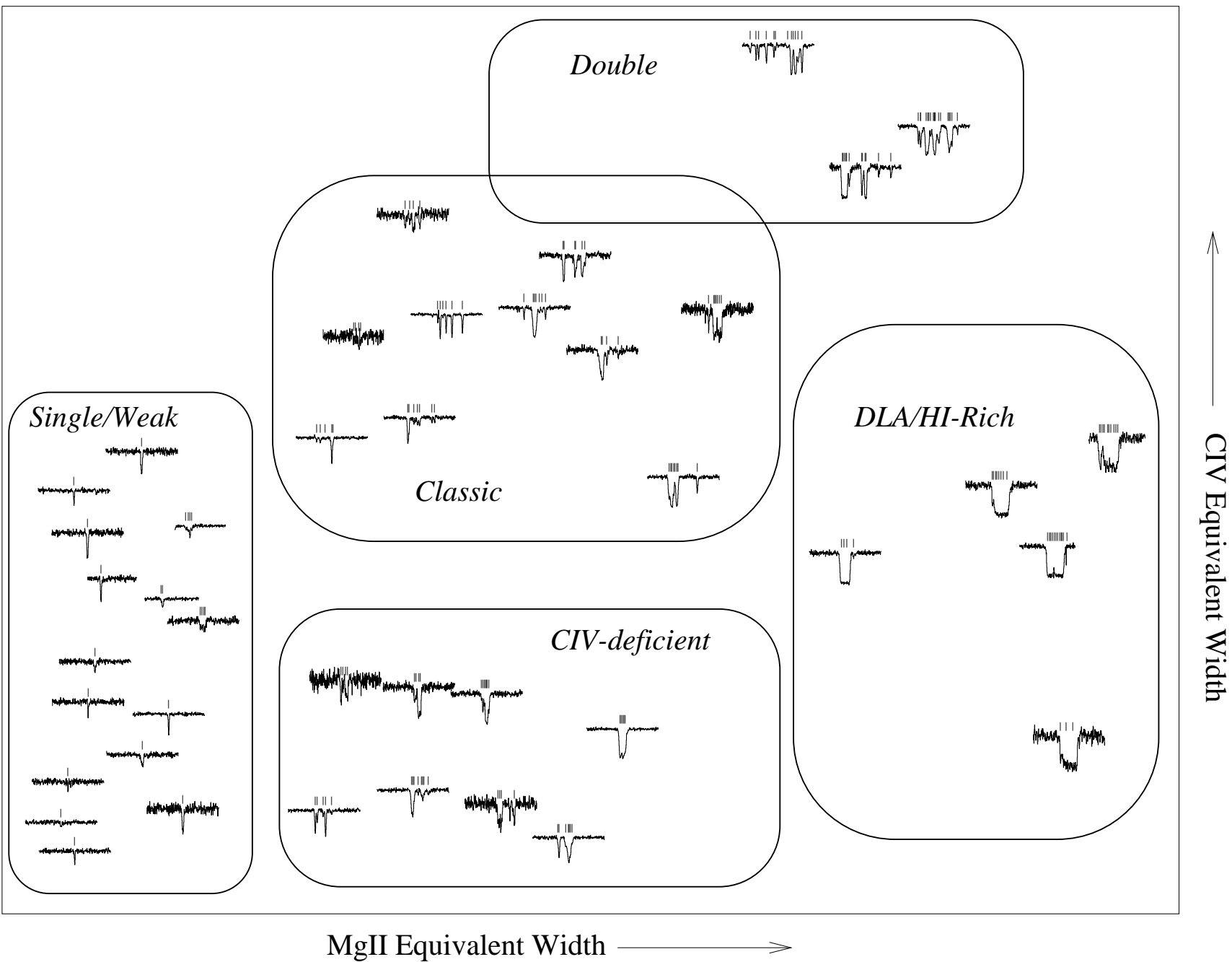}
\caption[fig13.ps]
{Taxonomy and the {\CIV}--{\MgII} kinematics
connection. The {\MgII} $\lambda 2796$ profile kinematics are
placed roughly by their {\CIV} and {\MgII} strengths on the
$W_{r}({\CIV})$--$W_{r}({\MgII})$ plane.  The locii of the taxonomic
classes are schematically represented by the outlines.
\label{fig:postage}}
\end{figure*}

\begin{figure*}[t]
\plotone{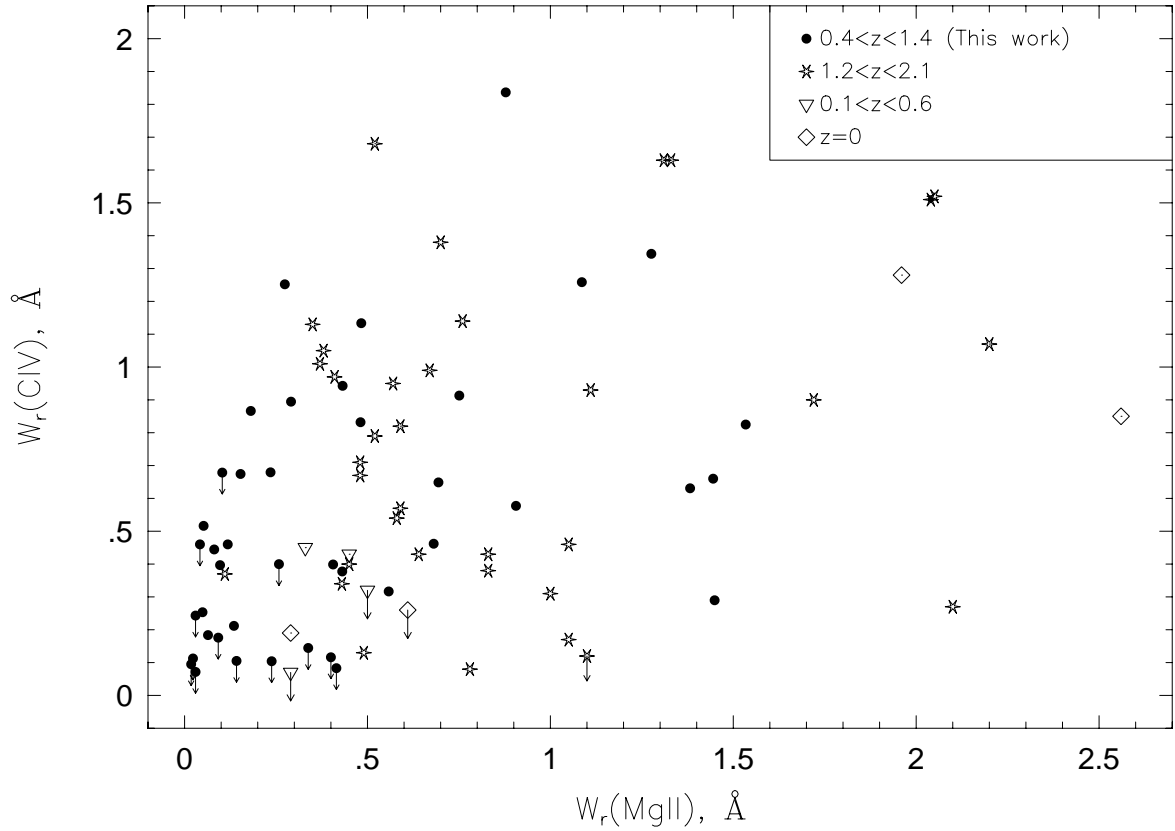}
\caption[fig14.ps]
{The $W_{r}({\CIV})$--$W_{r}({\MgII})$ plane.
Data from this study have point types the same as in
Figure~\ref{fig:k-means}.
Also plotted are:  higher redshift points ($1.2 \leq z \leq 2.2$) from
Bergeron \& Boiss\'{e} 1984, Boiss\'{e} \& Bergeron 1985, Lanzetta
\etal (1987) Steidel \& Sargent (1992) [six--pointed stars], lower
redshift points  ($0.1 \leq z \leq 0.6$) from Bergeron \etal (1994)
[downward pointing, open triangles], and $z\simeq 0$ data
from Bowen \etal (1996, 1996), and Bowen (private communication, 1999)
[open diamonds].
\label{fig:civmgii}}
\end{figure*}

\begin{figure*}[t]
\plotone{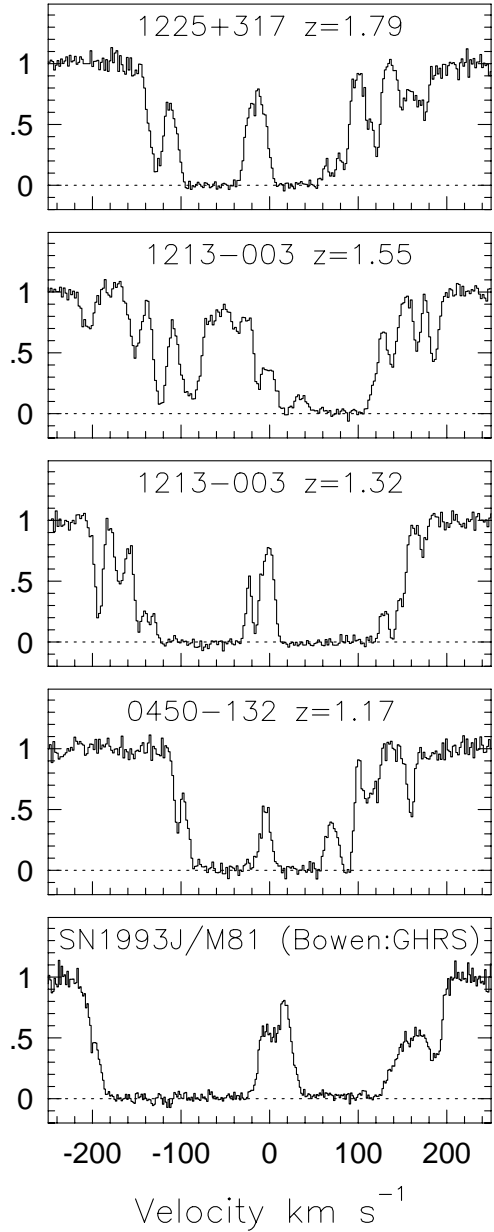}
\caption[fig15.ps]
{The {\MgII} $\lambda 2796$ profiles for four
higher redshift systems and for the $z\simeq 0$ SN 1993J, a line of
sight from the galaxy M81.  These systems all have equivalent widths
greater than $2$~{\AA}.  They occupy the extreme right portion of the 
{\CIV}--{\MgII} plane (Figure~\ref{fig:civmgii}) and are members of
the population of the most rapidly evolving {\MgII} absorbers.
\label{fig:highzMgII}}
\end{figure*}

The main results are summarized as follows:

1. 
The clustering analysis revealed that there is a wide range of
properties for {\MgII}--selected absorbers and that these can be 
categorized into three {\it main\/} types (Figure~\ref{fig:treecluster}).
Based upon strong $W_{r}({\Lya})$ and $W_{r}({\FeII})$, there is the
class of DLA/{\HI}--Rich {\MgII} absorbers.
For the remaining systems, there is a spread in the {\CIV} strengths
for a given $W_{r}({\MgII})$ that gives rise to `{\CIV}--weak'' and
``{\CIV}--strong'' systems.

2. 
The {\CIV}--weak class separates into the Single/Weak and the
{\CIV}--deficient classes (Figures~\ref{fig:treecluster} and
\ref{fig:k-means}).
Single/Weak systems are characterized by a single unresolved
{\MgII} absorption line with $W_{r}({\MgII}) \leq 0.15$~{\AA}, a
velocity spread of  $\omega _{v} \leq 6$~{\kms}, and a range of {\CIV}
strengths, but with $W_{r}({\CIV})$ no greater than $\sim 0.5$~{\AA}
(in our sample).
The {\MgII} strengths and kinematics are the dominant properties in
distinguishing them as a separate class.
{\CIV}--deficient systems have multiple Voigt profile components,
$N_{cl}>1$, a range of kinematic spreads, $15 \leq \omega_{v} \leq
45$~{\kms}, and $W_{r}({\CIV})$ no stronger than $0.4$~{\AA}.
The {\CIV} strength, and to a lesser degree the {\MgII} kinematics,
are the dominant properties distinguishing this as a 
{\CIV}--weak class.

3. 
The {\CIV}--strong class separates into the Classics and the Doubles
(Figures~\ref{fig:treecluster} and \ref{fig:k-means}).
Classic systems, like the {\CIV}--deficient systems,  have multiple
Voigt profile components, and a similar, but slightly larger, range of
kinematic spread.
The {\CIV} strengths, on the other hand, are greater than $0.4$~{\AA}.
Separate from the Classics are the Doubles, which are characterized by
roughly twice the number of Voigt profile components, $N_{cl}$, twice
the {\CIV} strength, and twice the {\MgII} kinematic spread, as
compared to the Classics.

4.
The {\MgII}, and {\Lya} strengths are tightly correlated with the
number of {\MgII} Voigt profile components, $N_{cl}$ (see
Figure~\ref{fig:nclouds}).
The tight correlation between $W_{r}({\Lya})$ and $N_{cl}$ implies
that the majority of the neutral hydrogen is arising in the phase
giving rise to the {\MgII} absorption.
In single cloud, $N_{cl}=1$, systems (the Single/Weak class),
$W_{r}({\Lya})$ ranges from $0.2$--$1.2$~{\AA} and $W_{r}({\CIV})$
ranges from less than $0.15$ to $0.7$~{\AA}, suggesting a range
of metallicities and/or ionization conditions in this class of object.

5. 
There is a highly significant correlation between
$W_{r}({\CIV})$ and $\omega_{v}$ (see Figure~\ref{fig:kinematics}; also
see \cite{letter}).
This correlation is driven by the five absorbers (two
Classics and three Doubles) with the largest $\omega _{v}$.  
For $\omega_{v} \leq 60$~{\kms} there is a $\sim 1$~{\AA} spread in
$W_{r}({\CIV})$, due to the {\CIV}--deficient systems and the
Single/Weak systems with larger $W_{r}({\CIV})/W_{r}({\MgII})$. 
More systems with $\omega _{v} \geq 80$~{\kms} are needed in order to
determine if all ``kinematically active'' {\MgII} absorbers have such
strong {\CIV}.

6.
Assuming the {\MgII} clouds are in photoionization equilibrium, we
showed that in many systems a substantial fraction of the {\CIV}
absorption is arising in a separate phase from the {\MgII} (see
Figure~\ref{fig:multi}).
We also showed that some fraction of the systems likely have {\SiIV}
in a separate phase from the {\MgII}.
The FOS profiles of the larger {\CIV} absorbers (Classics with large
{\CIV} and Doubles) are resolved due to the kinematic spread, or velocity
structure, of the {\CIV} phase.
A quantity expressing the degree to which the profiles are
resolved, $F_{r}$,  correlates at the 99.99\% confidence level with
the {\MgII} kinematic spread.
Furthermore, there is a clear impression that the asymmetries in the
resolved {\CIV} profiles trace the {\MgII} kinematics.
The above facts lead us to suggest that, often, some of the {\CIV}
arises in a physically distinct phase from the {\MgII} gas, but is
kinematically clustered with the {\MgII} clouds.

7. 
For our small sample of 16 galaxies, we find no significant
trends (greater than $3~\sigma$) of {\CIV} and {\Lya}, or their ratios
with each other and with {\MgII}, with host galaxy $B$ and $K$
luminosities, $B-K$ colors, and impact parameters.
DLA/{\HI}--Rich systems, which have systematically smaller impact
parameters and redder colors, significantly affect the outcome of
correlation tests.
The {\CIV} and {\Lya} absorption properties do not smoothly depend
upon global galaxy properties.
At $2.8~\sigma$, we find that $W_{r}({\SiIV})/W_{r}({\CIV})$
decreases with impact parameter, which could be interpreted as very
tentative evidence for a global galactocentric ionization/density
gradient (see Figure~\ref{fig:galimpacts}). 
The available data also show marginal trends for redder galaxies to
have lower ionization conditions (Figure~\ref{fig:galcolors}).
This is {\it suggestive\/} that {\CIV} phases may be systematically
smaller in the redder galaxies.

8. 
We found that the {\MgII} kinematics of a given absorber is related to
its location in the $W_{r}({\CIV})$--$W_{r}({\MgII})$ plane.
In a comparison of our intermediate redshift sample ($0.4 \leq z
\leq 1.4)$ with published low redshift ($0 \leq z \leq 0.6$) and high
redshift ($1.2 \leq z \leq 2.1$) samples, we found evidence for
redshift evolution in the $W_{r}({\CIV})$--$W_{r}({\MgII})$ plane.
This implies that other ``classes'' of absorbers are present at high
redshift, possibly including a ``Double DLA/{\HI}--Rich'' class.
There is also a group of ``super'' {\MgII} systems at $z\sim2$ that
have a large range of $W_{r}({\CIV})$ values.
We disussed a scenario in which the evolution of these strongest
{\MgII} absorbers could be due to the evolution of the number of galaxy
pairs and/or accreting LMC--like satellite galaxies.

\subsection{Further Ruminations}

In general, the observed range of {\MgII} kinematics and {\CIV}
absorption strengths could be due to a number of factors.  
These include global differences in gaseous conditions related to
environment (i.e.\ group or isolated galaxies) or galaxy morphology,
local variations dominated by passage through different parts of host
galaxies (e.g.\ halos, outer disks, spiral arms, etc.), or small scale
fluctuations in galactic interstellar and halo structures.
An additional important issue is whether characteristic absorption
properties might be tied to the evolutionary stage of the host galaxy.
This would imply that a given host galaxy may not always be of the
same ``absorber class'' throughout its evolution.

At higher redshifts ($z\geq2$), galaxy--galaxy interactions were more
common and no doubt played an important role in the kinematics and
multiphase ionization conditions in many {\MgII} absorbers (e.g.\
\cite{carilli92}; \cite{bowen93}; \cite{bowen95}).
A local analogue is seen in absorption associated with M61, a nearly
face--on, later--type  galaxy with enhanced star formation.
The {\MgII} velocity spread is $\sim 300$~{\kms} at an impact
parameter 21~kpc and there are two nearby galaxies (\cite{bowenm61}).

Since it is the kinematic {\it complexity\/} and spread and not the
total amount of the {\MgII} that appears to be the important property
for predicting the {\CIV} strength in $z\sim1$ galaxies, it is not
unreasonable to conjecture that sources of mechanical energy (i.e.\
active star formation) may be of central importance (see
\cite{guillemin}).
Extended multiphase halos are seen preferentially around late--type
spiral galaxies with signs of elevated star formatiopn (see
\cite{dahlem98} and references therein).
In fact, a basic scenario which might explain both the taxonomy of
{\MgII} absorbers (especially the Double and {\CIV}--deficient
systems) and the {\CIV}--{\MgII} kinematics connection, is one in
which star formation processes accelerate hot, ionized gas outward,
which subsequently cools and fragments into parcels of low ionization
gas (e.g.\ \cite{houck90}; \cite{li-ikeuchi92}; \cite{avillez99}).
These parcels would be observed as outlying kinematic components in
the {\MgII} profiles.

These points are somewhat suggestive of a causal connection between
evolution in the cosmic star formation rate and absorbing gas cross
sections, kinematics, and ionization conditions.
If true, we could hypothesize that scenarios of a cosmic mean history of
galaxy evolution inferred from the global record of star formation
history should be fully consistent with one based upon an ``absorption
line perspective''.

The global star formation rate appears to drop below $z\sim 1.0$
(\cite{lilly96}; \cite{madau}; however, see \cite{cowie}), implying a
reduction in the (infalling) gas reservoir for galaxies.
It could be that a majority of galaxies transform into
self--regulatory systems by $z\sim1$ and begin to evolve in a more
isolated fashion and in a direction dependent upon their ability to
continue forming stars.
Galaxies that are rich in gas and capable of forming large
numbers of molecular clouds in their interstellar media (i.e.\
late--type galaxies) would continue to form stars and exhibit
evolution, whereas those less capable would exhibit no discernable
evolution.
Such a scenario is consistent with the differential luminosity
evolution reported by Lilly \etal (1995\nocite{lilly95}) and is
not inconsistent with the tentative trends we see between galaxy color
and ionization conditions (also see \cite{guillemin}).

It is unfortunate that there currently are no {\MgII} absorption--line
samples for the highest redshifts ($z>2.2$), so an absorption line
perspective cannot yet be fully appreciated [however, see the
theoretical predictions of Rauch, Haehnelt, \& Steinmetz
(1997\nocite{rauch97})].


\acknowledgements
Support for this work was provided by the NSF (AST--9617185), and NASA
(NAG 5--6399 and AR--07983.01--96A) the latter from the Space
Telescope Science Institute, which is operated by AURA, Inc.,
under NASA contract NAS5--26555.
R. R. M. was supported by an NSF REU supplement.
B. T. J. acknowledges support from NOAO, which is operated by AURA,
Inc., under cooperative agreement with the NSF.
We thank Eric Feigelson for several very informative discussions on
the statistical treatment of data and as a valuable resource of
additional information.
Sandhya Rao and David Turnshek kindly shared unpublished data of
{\MgII}/DLA--selected galaxy properties, for which we are grateful.
We thank David Bowen for providing {\CIV} data prior to their
publication and for an electronic version of the {\MgII} $\lambda
2796$ profile from the GHRS spectrum of SN~1993J.
We also thank Bill Matthews for an enlightening discussion on the
properties of gas and star formation in elliptical galaxies and the
anonymous referee for a careful reading of the manuscript that lead to
significant improvements.


 
\newpage

\begingroup
\begin{table}
\tablenum{1}
\label{tab:lineprops}
\rotate[l]{\makebox[0.9\textheight][l]{\vbox{
\begin{center}
\vglue -0.7in
\begin{tabular}{cllcrlllcl}
\multicolumn{10}{c}{TABLE 1}\\
\multicolumn{10}{c}{\sc Absorber Properties}\\
\hline \hline
QSO & $z_{abs}$ & $W({\MgII})$ [{\AA}] & $\omega_v$ [km/s] & $N_{cl}$ &
         $W({\FeII})$ [{\AA}] & $W({\CIV})$ [{\AA}] & $W({\Lya}) $ [{\AA}] & Samples$^{c}$ & Class\\
\hline
\multicolumn{10}{c}{ }\\
$0002+051$ &  0.5915 &   $ 0.103\pm0.008$ &    $  5.0\pm  0.7$ &  1 &          $< 0.012$ &           $< 0.23$$^{a}$ &       \nodata & IC & Sngl/Wk \nl
 &  0.8514 &   $ 1.086\pm0.016$ &    $ 97.1\pm  4.6$ & 12 &   $ 0.419\pm0.022$ &     $ 1.26\pm0.06$ &     $ 2.47\pm0.08$ & CA & Double \nl
 &  0.8665 &   $ 0.023\pm0.008$ &    $  1.7\pm  0.5$ &  1 &          $< 0.010$ &     $< 0.11$$^{a}$ &     $ 0.81\pm0.10$ & IC & Sngl/Wk \nl
 &  0.9560 &   $ 0.052\pm0.007$ &    $  6.2\pm  1.0$ &  1 &          $< 0.005$ &     $ 0.52\pm0.04$ &     $ 0.85\pm0.07$ & CA,IC & Sngl/Wk \nl
$0058+019$ &  0.6127 &   $ 1.625\pm0.013$ &    $ 50.8\pm  0.3$ &  7 &   $ 1.274\pm0.037$ &            \nodata &     $ 6.77\pm0.40$ &  & DLA/{\HI} \nl
 &  0.7252 &   $ 0.253\pm0.012$ &    $  9.3\pm  0.5$ &  1 &          $< 0.034$ &            \nodata &     $ 0.46\pm0.13$ &  & Sngl/Wk \nl
$0117+213$ &  0.5764 &   $ 0.906\pm0.100$ &    $ 26.9\pm  1.0$ &  4 & $0.270\pm0.050$$^{a}$ &     $ 0.58\pm0.06$ &     $11.15\pm1.09$ & CA & DLA/{\HI} \nl
 &  0.7291$^{b}$ &   $ 0.238\pm0.009$ &    $ 44.2\pm  0.9$ &  5 &   $ 0.075\pm0.016$ &           $< 0.10$ &            \nodata & IC & {\CIV}--def \nl
 &  1.0480 &   $ 0.415\pm0.009$ &    $ 39.1\pm  0.7$ &  7 &   $ 0.065\pm0.010$ &           $< 0.08$ &     $ 1.93\pm0.04$ & CA,IC & {\CIV}--def \nl
 &  1.3250 &   $ 0.291\pm0.011$ &    $ 76.1\pm  1.5$ &  6 &   $ 0.026\pm0.009$ &     $ 0.89\pm0.03$ &     $ 1.51\pm0.03$ & CA,IC & Classic \nl
 &  1.3430$^{b}$ & $ 0.153\pm0.008$$^{a}$ &    $ 34.8\pm  2.9$ &  5 & $ 0.029\pm0.004$$^{a}$ &     $ 0.67\pm0.02$ &     $ 1.19\pm0.04$ & CA,IC & Classic \nl
$0454-220$ &  0.4744 &   $ 1.382\pm0.009$ &    $ 40.8\pm  1.0$ &  8 &   $ 0.975\pm0.029$ &     $ 0.63\pm0.03$ &     $ 3.74\pm0.05$ & CA & DLA/{\HI} \nl
 &  0.4833 &   $ 0.431\pm0.007$ &    $ 15.5\pm  0.5$ &  7 &   $ 0.162\pm0.043$ &     $ 0.38\pm0.11$ &     $ 1.56\pm0.03$ & CA,IC & {\CIV}--def \nl
$0454+039$ &  0.6428 &   $ 0.118\pm0.008$ &    $  4.8\pm  0.6$ &  1 &   $ 0.037\pm0.014$ &     $ 0.38\pm0.03$ &     $ 0.70\pm0.05$ & CA,IC & Sngl/Wk \nl
 &  0.8596 &   $ 1.445\pm0.014$ &    $ 42.8\pm  1.3$ & 12 &   $ 1.232\pm0.014$ &     $ 0.59\pm0.04$ &     $10.70\pm0.26$ & CA & DLA/{\HI} \nl
 &  0.9315 &   $ 0.042\pm0.005$ &    $  4.4\pm  0.9$ &  1 & $ 0.030\pm0.008$$^{a}$ &           $< 0.62$ &     $ 0.31\pm0.07$ & IC & Sngl/Wk \nl
 &  1.1532$^{b}$ &   $ 0.432\pm0.012$ &    $ 24.1\pm  1.7$ &  7 &   $ 0.084\pm0.015$ &     $ 0.94\pm0.06$ &     $ 1.56\pm0.01$ & CA,IC & Classic \nl
$0823-223$ &  0.7055 &   $ 0.092\pm0.007$ &    $ 11.9\pm  0.7$ &  1 &          $< 0.008$ &           $< 0.18$ &            \nodata & IC & Sngl/Wk \nl
 &  0.9110$^{b}$ &   $ 1.276\pm0.016$ &    $ 87.1\pm  1.1$ & 18 &   $ 0.416\pm0.026$ &     $ 1.34\pm0.06$ &     $ 2.68\pm0.07$ & CA,IC & Double \nl
$0958+551$ &  1.2113 &   $ 0.060\pm0.007$ &    $  3.4\pm  1.1$ &  1 &          $< 0.006$ &            \nodata &           $< 0.92$ & & Sngl/Wk \nl
 &  1.2724 &   $ 0.081\pm0.007$ &    $  4.9\pm  0.6$ &  1 &   $ 0.017\pm0.004$ &     $ 0.44\pm0.03$ &     $ 0.75\pm0.15$ & CA,IC & Sngl/Wk \nl
$1206+459$ &  0.9276 &   $ 0.878\pm0.016$ &    $116.2\pm  4.8$ & 12 &   $ 0.077\pm0.020$ &     $ 1.84\pm0.52$ &     $ 2.47\pm0.07$ & CA,IC & Double \nl
 &  0.9343 &   $ 0.049\pm0.005$ &    $  7.7\pm  0.8$ &  1 &          $< 0.004$ &     $ 0.25\pm0.05$ &     $ 0.47\pm0.07$ & CA,IC & Sngl/Wk \nl
$1241+176$ &  0.5505 &   $ 0.481\pm0.019$ &    $ 33.9\pm  4.1$ &  4 &   $ 0.236\pm0.048$ &     $ 0.83\pm0.07$ &            \nodata & IC & Classic \nl
 &  0.5584 &   $ 0.135\pm0.014$ &    $ 17.2\pm  1.1$ &  4 &          $< 0.012$ &     $ 0.21\pm0.06$ &            \nodata & IC & Sngl/Wk \nl
 &  0.8955 &   $ 0.018\pm0.005$ &    $  5.0\pm  1.2$ &  1 &          $< 0.005$ &           $< 0.10$ &     $ 0.45\pm0.05$ & IC & Sngl/Wk \nl
$1248+401$ &  0.7730 &   $ 0.694\pm0.009$ &    $ 53.6\pm  2.6$ &  8 &   $ 0.247\pm0.020$ &     $ 0.65\pm0.06$ &     $ 1.45\pm0.04$ & CA,IC & Classic \nl
 &  0.8546$^{b}$ &   $ 0.235\pm0.014$ &    $ 39.4\pm  3.9$ &  7 &   $ 0.031\pm0.007$ &     $ 0.68\pm0.06$ &     $ 1.46\pm0.29$ & CA,IC & Classic \nl
$1317+277$ &  0.6601$^{b}$ &   $ 0.338\pm0.011$ &    $ 36.1\pm  2.2$ &  8 &   $ 0.126\pm0.016$ &           $< 0.14$ &     $ 1.48\pm0.03$ & CA,IC & {\CIV}--def \nl
$1329+412$ &  0.5008 &   $ 0.258\pm0.035$ &    $ 25.6\pm  1.4$ &  4 &          $< 0.100$ &           $< 0.40$ &            \nodata & A & {\CIV}--def \nl
 &  0.8933 &   $ 0.400\pm0.023$ &    $ 55.0\pm  2.8$ &  4 &   $ 0.080\pm0.035$ &     $< 0.12$$^{a}$ &     $ 1.15\pm0.16$ & A,CA & {\CIV}--def \nl
 &  0.9739 &   $ 0.181\pm0.035$ &    $ 26.8\pm  2.7$ &  4 &          $< 0.028$ &     $ 0.87\pm0.05$ &     $ 1.15\pm0.23$ & CA & Classic \nl
 &  0.9984 &   $ 0.142\pm0.010$ &    $  6.7\pm  0.7$ &  1 &   $ 0.058\pm0.017$ &           $< 0.11$ &     $ 0.31\pm0.20$ & CA & Sngl/Wk \nl
$1354+195$ &  0.4566 &   $ 0.751\pm0.023$ &    $ 31.9\pm  1.7$ &  7 &   $ 0.149\pm0.088$ &     $ 0.91\pm0.04$ &     $ 1.72\pm0.08$ & CA & Classic \nl
 &  0.5215 &   $ 0.030\pm0.007$ &    $  6.6\pm  1.1$ &  1 &          $< 0.012$ &           $< 0.24$ &     $ 1.08\pm0.08$ & IC & Sngl/Wk \nl
$1622+238$ &  0.4720 &   $ 0.681\pm0.048$ &    $ 25.6\pm  3.1$ &  1 &          $< 0.118$ &     $ 0.46\pm0.14$ &     $ 1.27\pm0.27$ & A & Classic: \nl
 &  0.6561 &   $ 1.449\pm0.029$ &    $ 44.7\pm  2.3$ &  3 &   $ 1.015\pm0.050$ &     $ 0.29\pm0.09$ &     $ 9.42\pm0.34$ & A,CA & DLA/{\HI} \nl
 &  0.7971 &   $ 0.274\pm0.029$ &    $ 45.2\pm  2.8$ &  4 &          $< 0.092$ &     $ 1.25\pm0.06$ &     $ 1.38\pm0.14$ & CA & Classic \nl
 &  0.8913 &   $ 1.534\pm0.025$ &    $ 50.7\pm  2.0$ & 10 &   $ 1.080\pm0.421$ &     $ 0.82\pm0.05$ &     $ 2.98\pm0.10$ & CA & DLA/{\HI} \nl
$1634+706$ &  0.8182 &   $ 0.030\pm0.018$ &    $  3.1\pm  1.6$ &  1 &          $< 0.008$ &           $< 0.07$ &            \nodata & IC & Sngl/Wk \nl
 &  0.9056 &   $ 0.064\pm0.004$ &    $  4.1\pm  0.6$ &  1 &          $< 0.005$ &     $ 0.18\pm0.02$ &     $ 0.49\pm0.03$ & CA,IC & Sngl/Wk \nl
 &  0.9902 &   $ 0.558\pm0.005$ &    $ 17.8\pm  0.2$ &  5 &   $ 0.127\pm0.011$ &     $ 0.32\pm0.02$ &     $ 1.09\pm0.03$ & CA,IC & {\CIV}--def \nl
 &  1.0414 &   $ 0.097\pm0.008$ &    $ 17.2\pm  1.1$ &  4 &          $< 0.038$ &     $ 0.40\pm0.02$ &     $ 1.42\pm0.01$ & CA,IC & Sngl/Wk \nl
$2128-123$ &  0.4297 &   $ 0.406\pm0.014$ &    $ 15.8\pm  0.4$ &  4 &    $< 0.260$$^{a}$ &     $ 0.40\pm0.04$ &     $ 2.92\pm0.08$ & CA & {\CIV}--def \nl
$2145+067$ &  0.7908 &   $ 0.483\pm0.015$ &    $ 75.1\pm  1.1$ &  6 &   $ 0.041\pm0.014$ &     $ 1.13\pm0.08$ &     $ 1.22\pm0.04$ & CA,IC & Classic \nl
\multicolumn{10}{c}{ }\\
\hline
\multicolumn{10}{l}{$^{a}$ Not ``flag'' transition; see system notes (\S3)}\\
\multicolumn{10}{l}{$^{b}$ Measurement of $\omega _{v}$ was censored
(see text).  Uncensored values are 46.8~{\kms} ($0117+213$, $z=0.7291$),
45.1~{\kms} ($0117+213$, $z=1.3430$),}\\
\multicolumn{10}{l}{$^{c}$ Sample A: systems with $W({\MgII})$
detection limits greater then $0.3$~{\AA}. Sample CA: systems used for
multivariate Cluster Analysis}.  \\
\multicolumn{10}{l}{Sample IC: systems used for Ionization Condistions
analysis.}\\
\end{tabular}
\end{center}
}}}
\end{table}
\endgroup


\newpage
 
\begin{table}
\tablenum{2}
\label{tab:galprops}
 
\begin{center}
\begin{tabular}{llrrrrl}
\multicolumn{7}{c}{TABLE 2} \\
\multicolumn{7}{c}{\sc Galaxy Properties} \\
\hline \hline
QSO         &  $z_{\rm abs}$ & $M_B$   & $M_K$   & $B-K$   & $D~h^{-1}$ & Reference\\ 
            &                &         &         &         & (kpc) & \\
\hline 
$0002+051$  &   $0.5915$     & $-21.1$ & $-25.2$ & $4.1$   & $23.8$  & Churchill \etal (1996) \\
            &   $0.8514$     & $-21.2$ & $-24.1$ & $2.9$   & $18.6$  & Churchill \etal (1996) \\
$0058+051$  &   $0.6127^{a}$     & $-19.6$ & $-22.7$ & $3.1$   & $19.3$  & This paper \\ 
$0117+213$  &   $0.5764$     & $-21.9$ & $-25.9$ & $4.0$   & $5.1$   & Churchill \etal (1996) \\
            &   $0.7291$     & $-22.3$ & $-26.3$ & $4.0$   & $36.0$  & Churchill \etal (1996) \\
$0454+039$  &   $0.8596$     & $-19.5$ & $>-23.4$ & $<3.9$ & $10.6$  & Churchill \etal (1996) \\
            & $1.1532^{b}$   & $-21.0$ & $-23.8$ & $2.8$   & $40.1$  & This paper \\
$1241+176$  &   $0.5505$     & $-20.5$ & $-23.9$ & $3.4$   & $13.8$  & Churchill \etal (1996) \\
$1248+401$  &   $0.7730$     & $-20.3$ & $-23.5$ & $3.2$   & $23.3$  & Churchill \etal (1996) \\
$1317+277$  &   $0.6601$     & $-21.9$ & $-25.7$ & $3.8$   & $37.6$  & Churchill \etal (1996) \\  
$1354+195$  &   $0.4566$     & $-20.7$ & $-23.8$ & $3.0$   & $26.2$  & This paper \\
$1622+238$  &   $0.4720$     & $-19.5$ & $-22.2$ & $2.7$   & $22.2$  & Steidel \etal (1997) \\
            &   $0.7971$     & $-21.5$ & $-24.6$ & $3.0$   & $46.5$  & Churchill \etal (1996), Steidel \etal (1997) \\
            &   $0.8913$     & $-20.6$ & $-23.7$ & $3.1$   & $15.0$  & Churchill \etal (1996), Steidel \etal (1997) \\
$2128-123$  &   $0.4297$     & $-21.4$ & $-24.7$ & $3.3$   & $31.6$  & This paper \\
$2145+067$  &   $0.7908$     & $-21.6$ & $-24.5$ & $3.0$   & $29.3$  & This paper \\
\hline
\multicolumn{7}{l}{$^{a}$ Recent analysis of a WFPC2/{\it HST\/} image
revealed a previously unseen galaxy within 1{\arcsec} of the QSO. This} \\
\multicolumn{7}{l}{galaxy identification should be viewed with caution.} \\
\multicolumn{7}{l}{$^{b}$ $M_B$ extrapolated from $\Re$--band image using 
an Im spectroscopic template consistent with the $R-K$ color.}
\end{tabular}
\end{center}
\end{table}
 

\newpage

\begin{table}
\tablenum{3}
\label{tab:civ-literature}
\begin{center}
\begin{tabular}{lcccc}
\multicolumn{5}{c}{TABLE 3}\\
\multicolumn{5}{c}{\sc {\MgII} and {\CIV} Equivalent Widths$^{a}$ from
the Literature}\\
\hline \hline
QSO & $z_{abs}$ & $W({\MgII})$ [{\AA}] & $W({\CIV})$ [{\AA}] &
References \\
\hline
Mrk~205    & $0.004$  & $0.29$ & $0.19$  & BB93 \\
$1219+047$ & $0.005$  & $1.96$ & $1.28$  & BBP96 \\
$0955+326$ & $0.005$  & $2.56$ & $0.85$  & B99,J98 \\
$1543+489$ & $0.075$  & $0.61$ & $<0.26$ & BBP95 \\
$1137+660$ & $0.1164$ & $0.50$ & $<0.32$ & B94 \\
$1704+608$ & $0.2216$ & $0.45$ & $0.43$  & B94 \\
$1317+277$ & $0.2891$ & $0.33$ & $0.45$  & B94 \\
$1634+706$ & $0.6694$ & $0.29$ & $<0.07$ & B94 \\
$0454+034$ & $1.1535$ & $0.57$ & $0.95$  & SS92 \\
$1101-264$ & $1.1875$ & $0.41$ & $0.97$  & BB85 \\
$1101-264$ & $1.2030$ & $0.59$ & $0.57$  & BB85 \\
$1247+265$ & $1.2233$ & $0.48$ & $0.67$  & SS92 \\
$1602-002$ & $1.3245$ & $0.64$ & $<0.43$ & LTW87 \\
$0226-035$ & $1.3277$ & $0.83$ & $0.43$  & SS92 \\
$1421+122$ & $1.3605$ & $0.43$ & $0.34$  & LTW87 \\
$0237-232$ & $1.3647$ & $2.05$ & $1.52$  & SS92 \\
$0957+561$ & $1.3901$ & $2.10$ & $0.27$  & BB84 \\
$0058+016$ & $1.4641$ & $0.45$ & $0.40$  & SS92 \\
$0002-422$ & $1.5413$ & $0.48$ & $0.71$  & LTW87 \\
$0424-131$ & $1.5615$ & $0.38$ & $1.05$  & SS92 \\
$1329+412$ & $1.6012$ & $0.70$ & $1.38$  & SS92 \\
$1017+276$ & $1.6081$ & $0.11$ & $0.37$  & SS92 \\
$0237-232$ & $1.6363$ & $0.78$ & $0.08$  & SS92 \\
$0421+016$ & $1.6375$ & $0.35$ & $1.13$  & SS92 \\
$1246-054$ & $1.6466$ & $0.52$ & $0.79$  & SS92 \\
$0237-232$ & $1.6583$ & $0.76$ & $1.14$  & SS92 \\
$0237-232$ & $1.6732$ & $1.31$ & $1.63$  & SS92 \\
$1311-270$ & $1.6860$ & $0.83$ & $<0.38$ & LTW87 \\
$0958+551$ & $1.7320$ & $0.37$ & $1.01$  & SS92 \\
$0854+191^{b}$ & $1.7342$ & $4.14$ & $1.14$  & SS92 \\
$1017+276$ & $1.7953$ & $2.04$ & $1.51$  & SS92 \\
$1331+170$ & $1.7860$ & $1.05$ & $0.17$  & SS92 \\
$1331+170$ & $1.7766$ & $1.33$ & $1.63$  & SS92 \\
$0100+130$ & $1.7971$ & $1.11$ & $0.93$  & SS92 \\
$1704+710$ & $1.8107$ & $0.58$ & $0.54$  & SS92 \\
$1329+412$ & $1.8355$ & $0.49$ & $0.13$  & SS92 \\
$1101-264$ & $1.8386$ & $1.00$ & $0.31$  & LTW87 \\
$1228+077$ & $1.8966$ & $1.72$ & $0.90$  & LTW87 \\
$1435+635$ & $1.9235$ & $1.05$ & $0.46$  & SS92 \\
$0551-366^{b}$ & $1.9613$ & $5.80$ & $2.13$  & LTW87 \\
$0119-044$ & $1.9638$ & $0.52$ & $1.68$  & SS92 \\
$0013-003^{b}$ & $1.9713$ & $7.37$ & $2.19$  & SS92 \\
$2126-158$ & $2.0219$ & $0.67$ & $0.99$  & LTW87 \\
$0013-003^{b}$ & $2.0290$ & $5.75$ & $0.63$  & SS92 \\
$0348+061$ & $2.0237$ & $0.59$ & $0.82$  & SS92 \\
$0424-131$ & $2.0344$ & $1.10$ & $<0.12$ & SS92 \\ 
$0450-131$ & $2.0667$ & $2.20$ & $1.07$  & SS92 \\
\hline
\multicolumn{5}{l}{$^{a}$ Equivalent widths are rest frame.} \\
\multicolumn{5}{l}{$^{b}$ Does not appear on Figure~\ref{fig:civmgii}.} \\
\multicolumn{5}{l}{References:--- (BB84) Bergeron \& Boiss\'{e} 1984;
(BB85) Boiss\'{e} \& Bergeron 1985;} \\
\multicolumn{5}{l}{(LTW87) Lanzetta \etal 1987; (BB93) Bowen \&
Blades 1993; (B94) Bergeron} \\
\multicolumn{5}{l}{\etal 1994; (BBP95) Bowen \etal 1995; (BBP96) Bowen
\etal 1996; (SS92)} \\
\multicolumn{5}{l}{Steidel \& Sargent 1992; (J98) Jannuzi \etal 1998;
(B99) Bowen 1999.} \\
\end{tabular} 
\end{center}
\end{table}
 


\clearpage

\clearpage

\clearpage

\clearpage

\clearpage

\clearpage

\clearpage

\clearpage

\clearpage

\clearpage

\clearpage

\clearpage

\clearpage

\clearpage

\clearpage

\end{document}